%% file: main.tex
\documentclass[runningheads]{llncs}
\usepackage[T1]{fontenc}
\usepackage{graphicx}
\usepackage{setspace}
\usepackage{xcolor}
\usepackage{tcolorbox}
\tcbuselibrary{theorems,skins}
\usepackage{color}
\usepackage{ulem}
\usepackage{amsfonts}
\usepackage{subcaption}
\usepackage{ascmac}
\usepackage{fancybox}
\usepackage{colortbl}
\usepackage{multirow}
\usepackage{fancyvrb}
\usepackage{wrapfig}
\usepackage{tabularray}
\definecolor{mygray}{gray}{0.9}
\newcommand{\react}{\mathrel{\texttt{:-}}}
\newcommand{\equals}{\mathop{\texttt{=}}}

\newcommand{\reducesR}[1]{\xrightarrow{\,#1\,}}
\newcommand{\mem}[1]{\texttt{\{}#1\texttt{\}}}
\newcommand{\mema}[1]{\{ #1\}}
\newcommand{\Q}{{\cal Q}}
\newcommand{\card}[4]{{#1\texttt{<}#2\texttt{,}#3\texttt{>}#4}} 
\newcommand{\carda}[4]{{#1\langle #2,#3\rangle #4}} 
\newcommand{\num}[3]{{#1\texttt{<}#2\texttt{>}#3}} 
\newcommand{\numa}[3]{{#1\langle #2\rangle #3}} 
\newcommand{\func}{\textit{simp}}
\newcommand{\context}{\textit{cxt}}
\newcommand{\negation}{\textit{neg}}
\newcommand{\pc}{\texttt{\$}}
\newcommand{\self}{\gamma}
\let\caret=\textasciicircum 
\newcommand{\itnameid}[1]{#1^{id}} 
\newcommand{\ttnameid}[1]{$\texttt{#1}^{id}$} 
\newcommand{\itnamecxt}[1]{#1^{ct}} 
\newcommand{\ttnamecxt}[1]{$\texttt{#1}^{ct}$} 
\newcommand{\step}[2]{\vspace{10pt}\noindent\textit{#1} #2}
\newcommand{\narrowdots}{.\kern1pt.\kern1pt.\kern1pt}
\newcommand{\newsentence}[1]{{\color{black}#1}}
\newcommand{\mishinasentence}[1]{{\color{black}#1}}
\newcommand{\todo}[1]{}

\begin{document}
\title{Introducing Quantification \\ into a Hierarchical Graph
  Rewriting Language\thanks{This is an extended version (with Appendix)
    of the paper
    presented at the 34th International Symposium on Logic-Based
    Program Synthesis and Transformation (LOPSTR 2024), Milano, Italy,
    September 2024, LNCS 14919, Springer-Verlag, pp.~220--239,
    https://doi.org/10.1007/978-3-031-71294-4\_13.}}
\titlerunning{Introducing Quantification into LMNtal}
%
\author{Haruto Mishina\inst{1} \and
Kazunori Ueda\inst{2}\orcidID{0000-0002-3424-1844}}
\authorrunning{H. Mishina and K.Ueda}
%
\institute{Waseda University, Tokyo 169-8555, Japan\\
\email{\{mishina,ueda\}@ueda.info.waseda.ac.jp}}
\maketitle              
\begin{abstract}

LMNtal is a programming and modeling language based on hierarchical
graph rewriting that uses logical variables to represent connectivity
and membranes to represent hierarchy.  On the theoretical side, it
allows logical interpretation based on intuitionistic linear logic;
on the practical side, its full-fledged implementation supports
a graph-based parallel model checker and has been used to
model diverse applications including various computational models.
This paper discuss how we extend LMNtal to QLMNtal (LMNtal with
Quantification) to further enhance the usefulness of hierarchical
graph rewriting for high-level modeling by introducing quantifiers
into rewriting as well as matching.
%
Those quantifiers allows us to express universal quantification,
cardinality and non-existence in an integrated manner.
Unlike other attempts to introduce quantifiers into graph rewriting,
QLMNtal has term-based syntax, 
whose
semantics is smoothly
integrated into the small-step semantics of the base language LMNtal.
The proposed constructs allow combined and nested use of
quantifiers within individual rewrite rules.

\keywords{quantification \and graph rewriting language \and language design.}
\end{abstract}
\input{tex/sec1}

\input{tex/sec2}

\input{tex/sec3}

\input{tex/sec4}

\input{tex/sec5}

\input{tex/sec6}

\input{tex/sec7}

\input{tex/ack}

\bibliographystyle{splncs04}
\bibliography{main}

\input{tex/appendix}

\end{document}

%% file: tex/sec1.tex
\section{Introduction}
\label{sec:introduction}

\subsection{Background and Objectives}

Graph rewriting languages express computation 
as successive and possibly concurrent updating of graph structures.
Graphs are highly general means for modeling diverse structures in the
real world. 
They also generalize algebraic data types such as trees and lists
in programming, and designing high-level programming languages for handling
graphs safely and clearly is an important challenge.

A specific challenge towards an expressive graph rewriting language,
both in theory and in practice, is the ability to handle the
``quantities''---broadly construed---of graph elements.
They include existential and
universal quantification, 
non-existence, and cardinality.
Existing 
tools for graph rewriting
have proposed various methods including the combination of rewrite
rules and the introduction of control
structures (see Section~\ref{sec:related}
for related work), but
how to provide those features in the standard setting of programming
languages, i.e., inductively defined syntax and structural operational
semantics, has been an open question.


LMNtal \cite{LMNtal} is a programming and modeling language that handles
hierarchical undirected graphs.  Its distinguished features include:
\begin{itemize}
\item It is based on inductively defined syntax and small-step
  semantics,
  with structural
  congruence (as with process calculi) to characterize its data
  structures which are often referred to as \textit{port graphs} 
  (as opposed
  to standard graphs in graph theory).  Links (which are unlabeled edges)
  interconnect two ports of atoms (which are labeled nodes).
\item It features \textit{membranes} as a means of hierarchization (of
  nodes), representation of first-class multisets, and localization of
  rewriting.
\item The de facto standard runtime system, SLIM \cite{Gocho2011},
  provides an LTL
  model checker as well as (don't-care non-deterministic) concurrent
  rewriting.
\end{itemize}

Historically, LMNtal was born as an attempt to unify constraint-based
concurrency \cite{FGCS-SCP} and Constraint Handling Rules \cite{CHR}, the
two notable extensions to (concurrent) logic programming. This attempt
then resulted in the unified handling of symbols representing programs
\newsentence{and} those representing data, which are interconnected by logical
variables that could be interpreted as graph edges.
LMNtal programs allow interpretation based on intuitionistic linear
logic \cite{LMNtal}.

Unlike many other tools for graph rewriting that provide control
structures (like sequencing), computation of LMNtal is basically
controlled by subgraph matching which can be regarded essentially
as existential quantification and which 
acts as a
\textit{synchronization mechanism}.  
The
implementation of LMNtal supports constructs for
handling a limited form of negative information 
including the non-equality checking 
of node labels, but 
constructs for handling various sorts of quantities 
has been an open issue, which is the subject 
of this paper.

\subsection{Contributions}

The objective of this work is to introduce constructs for specifying
quantities into a graph rewriting language
and to formulate the syntax and semantics of the extended language.
The 
contributions of our approach are summarized below.

\medskip\noindent
\textbf{Introducing cardinality, non-existence and universal quantification.}
Pattern matching in LMNtal is to find a subgraph which is structurally
congruent to the LHS (left-hand side) of a rewrite rule 
and can be viewed as existential quantification.  We first introduce
constructs for specifying the quantities of graph elements.
\textit{Cardinality quantification} specifies the
minimum and the maximum numbers of specified subgraph occurrences
in the target graph and 
\newsentence{rewrites} them in a single step.
\textit{Non-existence quantification} ensures the
non-existence of a specified subgraph in the target graph.
\textit{Universal quantification} finds all (non-overlapping)
subgraphs
and rewrite them in a single step.

\medskip\noindent
\textbf{Relating different \newsentence{quantifications}.}
One of the technical challenges is to introduce the above 
quantification systematically in such a way
that different constructs are related to each other.
We identify cardinality and non-existence as two
basic quantification and universal quantification as a derived
construct.  Also, we show that \textit{labelling} of quantification
plays a key role in controlling the (in)dependence of quantification.

\medskip\noindent
\textbf{Abstract state space.}
Although various forms of quantification could be programmed using
multiple rewrite rules and/or built-in control structures, they
%
\newsentence{complicate} the state space of model checking.
Expressive quantification within individual rules
will keep the state space at the right level of abstraction.

\medskip\noindent
\textbf{Combination and nesting of quantification.}
%
Thanks to the approach
based on structural operational semantics, the semantics of combined
and nested use of quantification can be given in a systematic manner.
We show that the proposed framework can express typical use cases of
nested quantification (Section~\ref{sec:more_examples}).


\medskip
The rest of this paper is organized as follows.
Section \ref{sec:lmntal} describes LMNtal, the base language of this study. 
Section \ref{sec:introductoryexamples}
introduces QLMNtal's
functionalities using simple examples.
The syntax of QLMNtal is given Section \ref{sec:syntax} and the
operational semantics in Section \ref{sec:semantics}.
Section \ref{sec:more_examples} describes two examples showing the
expressive power of the constructs.
Section \ref{sec:related} discusses related work and our 
future work.

%% file: tex/sec2.tex
\section{LMNtal}
\label{sec:lmntal}

This section outlines (a fragment of) LMNtal \cite{LMNtal} 
as the 
basis of
QLMNtal.  Due to its background as a concurrent language, 
its main
construct is called a \textit{process},
\newsentence{whose structure formed by atoms, links and membranes will
evolve by another construct called \textit{(rewrite) rules}.}
The original definition
\cite{LMNtal} handles rewrite rules as part of a process so that they
can be placed in membranes to express local rewriting 
inside them,
but
here we
handle only global rewrite rules and
separate them from processes as in \cite{Access}.

\subsection{Syntax of LMNtal}\label{sec:syntax_LMNtal}

An LMNtal program discussed in this paper is the pair of a
\textit{process} and a set of \textit{(rewrite) rules} defined as in
Fig.~\ref{fig:Syntax of LMNtal},
where $p$ stands for a name starting with a lowercase letter or a
number, and $X_i$ stands for a name starting with an uppercase
letter. 
Processes and rules are subject to the Link Condition:
\begin{definition}[Link Condition]
Each link name can occur at most twice in a process, 
and each link name in a rule must occur exactly twice in the rule.\qed
\end{definition}

\begin{figure}[t]
	\centering
	\begin{tabular}{rp{10pt}cll}
	\hline\\[-6pt]
(process) & $P$ & $::=$ & $\mathbf{0}$ & (null) \\
	  & & $|$ & $p(X_1,\dots,X_m)$\quad$(m\ge 0)$ & (atom) \\
	  & & $|$ & $P,P$ & (parallel composition) \\
	  & & $|$ & $\mem{P}$ & (membrane) \\[3pt]
(rule)    & $R$ & $::=$ & $T\react T$ & \\[3pt]
(template)& $T$ & $::=$ & $\mathbf{0}$ & (null) \\
	  & & $|$ & $p(X_1,\dots,X_m)$\quad$(m\ge 0)$ & (atom) \\
	  & & $|$ & $T,T$ & (parallel composition) \\
	  & & $|$ & $\mem{T}$ & (membrane) \\
	  & & $|$ & $\texttt{\$}p$ & (process context) \\[3pt]
		\hline
	\end{tabular}
	\caption{Syntax of LMNtal}
	\label{fig:Syntax of LMNtal}
\vspace*{-10pt}
\end{figure}

An $m$-ary \textit{atom} $p(X_1,\dots,X_m)$ represents a node 
\newsentence{(labelled by the name $p$)}
of a port graph
with $m$ totally ordered ports to which the links $X_1,\dots,X_m$ are
connected.  

A \textit{link} represents an undirected edge.  A link occurring twice in a
process is called a \textit{local link}, while a link occurring once in a
process is called a \textit{free link}.

\textit{Parallel composition}
glues
two processes to build a
larger process.
Note 
that, if each of 
$P_1$ and $P_2$ has a free link with the same name, it becomes a local
link in $(P_1, P_2)$.
A 
reader may notice that the Link Condition may not
  always allow us to form 
$(P_1, P_2)$ from
  two graphs $P_1$ and $P_2$ each satisfying the Link Condition.
  How LMNtal handles this point will be discussed in
  Section~\ref{sec:SC_LMNtal}.

A special binary atom, called a \textit{connector} $\texttt{=}(X,Y)$,
also written as $X\equals Y$,
fuses (or glues) two links $X$ and $Y$.

A \textit{membrane} acts as a means to group atoms and other
membranes and forms a hierarchical structure.  A link may cross
membranes to connect two atoms belonging to two different places of
the hierarchical membrane structure.

A \textit{(rewrite) rule} consists of two templates,
called the \textit{head} (the left-hand side)
and the \textit{body} (the right-hand side) of the rule, respectively.
%

A \textit{process context} $\texttt{\$}p$ in the head of a rule
acts as the \textit{wildcard} of a membrane
(which contains the $\texttt{\$}p$ at its top level; 
\newsentence{i.e., not inside another membrane contained by the
  aforementioned membrane)} in pattern matching.
A process context \newsentence{$\texttt{\$}p$ with the name $p$} may appear
(i) \newsentence{at most}
once in the head of a rule as the sole process context of some
membrane (i.e., a membrane can contain at most one process context at
its top level), and 
(ii) \newsentence{if it does, it must appear}
exactly once in the body of the rule.


Note that a process context in (full) LMNtal \cite{LMNtal} may
optionally specify the list of free links of the process context, but
we omit the description \newsentence{of this feature}
because it is not used in the examples of the present paper.
\newsentence{We just note that a process context of the form
$\texttt{\$}p$ may match a process with any number of free links.}


A term representing a process is subject to Structural Congruence
defined in Section~\ref{sec:SC_LMNtal},
which then stands for an undirected
multigraph, i.e., a graph that allows multi-edges and self-loops.

\begin{figure}[t]
	\begin{minipage}[b]{0.32\linewidth}
		\begin{screen}[4]
			\small\texttt{a(V,W,Z),b(X,Y,Z)}\\[14pt]
			\centering
			\includegraphics[width=3cm]{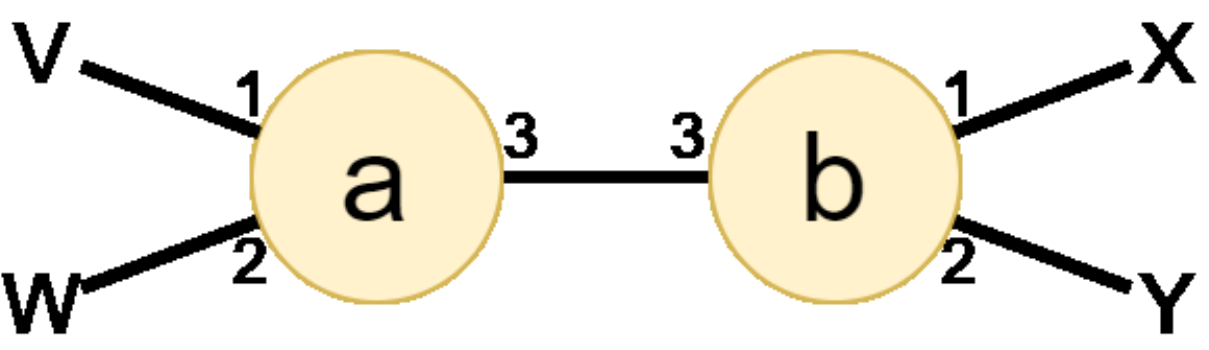}
			\vspace{6pt}
		\end{screen}
		\caption{An LMNtal process and its visualization}
		\label{fig:LMNtal Process}
	\end{minipage}\hfil
	\begin{minipage}[b]{0.25\linewidth}
		\centering
		\includegraphics[width=2.5cm]{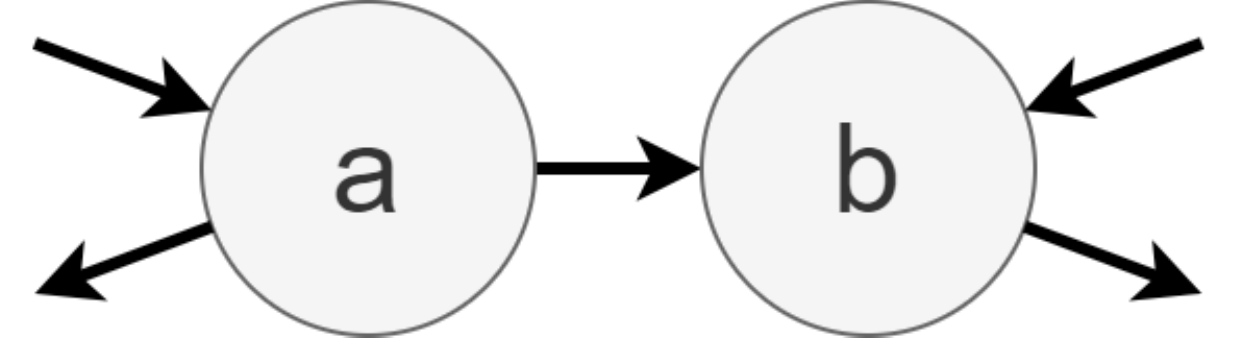}
		\vspace{20pt}
		\caption{A standard directed graph}
		\label{fig:Unordered Directed Graph}
	\end{minipage}\hfil
	\begin{minipage}[b]{0.36\linewidth}
		\begin{screen}[4]
			\small\texttt{\{a,t(V),s(W),s(Z)\},}\\
			\small\texttt{\{b,t(X),s(Y),t(Z)\}}\\[6pt]
			\centering
			\includegraphics[width=3.5cm]{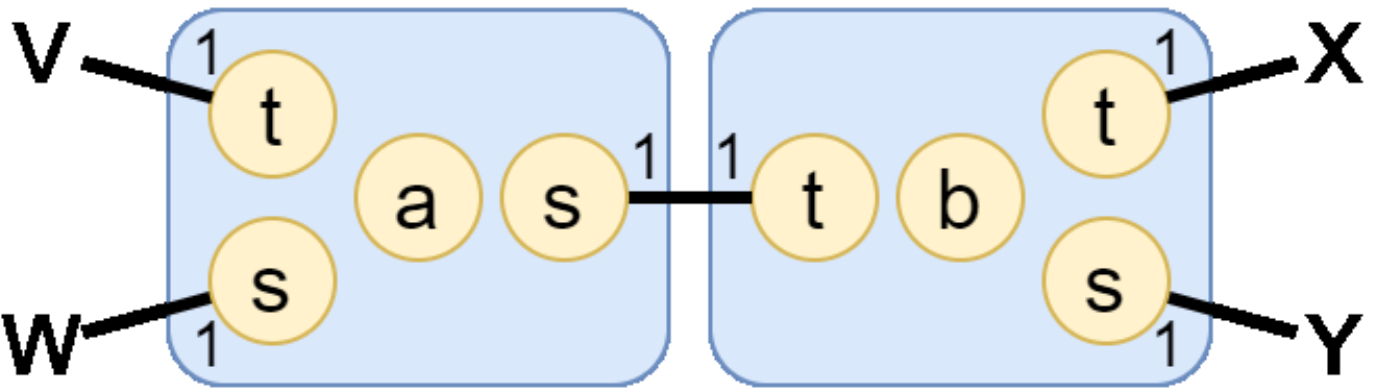}
		\end{screen}
		\caption{An LMNtal representation of
                  Fig.~\ref{fig:Unordered Directed Graph}} 
		\label{fig:LMNtal Representation}
	\end{minipage}
\end{figure}

Figure~\ref{fig:LMNtal Process} shows a simple 
process
and its visualization as a port graph, in which the
local link name \newsentence{\texttt{Z}}
is insignificant and not shown.
Unlike LMNtal atoms, nodes of standard graphs in graph theory 
come with unordered edges and the graphs are often directed,
as in Fig.~\ref{fig:Unordered Directed Graph}.
Those standard graphs can be represented in LMNtal
as in Fig.~\ref{fig:LMNtal Representation}, where
each node is represented by a membrane containing an atom representing
the node label, 
and each edge is represented by 
a
link connecting a unary
\texttt{s}(ource) atom and a unary \texttt{t}(arget) atom.

\subsection{Semantics of LMNtal}\label{sec:semantics_LMNtal}

The semantics of LMNtal consists of \textit{structural congruence} and
\newsentence{a}
\textit{reduction relation}.

\subsubsection{Structural Congruence}\label{sec:SC_LMNtal}

The syntax defined above does not yet characterize LMNtal graphs
because the port graph of Fig.~\ref{fig:LMNtal Process} allows
other syntactic representations, e.g., 
\verb|b(X,Y,U),a(V,W,U)|.

Figure~\ref{fig:Structural congruence on LMNtal processes} 
defines an equivalence relation,
called
\textit{structural congruence}, to absorb the syntactic variations.
$P[Y/X]$ stands for the renaming of link $X$ in process $P$ to $Y$. 
The rules apply only when each process satisfies the Link Condition.
The rules are as in \cite{LMNtal} and readers familiar with
LMNtal may skip the details. 

\begin{figure}[tbp]
	\centering
	\begin{tabular}{rr@{~~}c@{~~}l@{~~}l}
		\hline\\[-6pt]
		(E1) & $0,P$ & $\equiv$ & $P$ \\[1pt]
 		(E2) & $P,Q$ & $\equiv$ & $Q,P$ \\[1pt]
		(E3) & $P,(Q,R)$ & $\equiv$ & $(P,Q),R$ \\[1pt]
		(E4) & $P$ & $\equiv$ & $P[Y/X]$ & if $X$ is a local link of $P$ \\[1pt]
		(E5) & $P\equiv P'$ & $\Rightarrow$ & $P,Q\equiv P',Q$ \\[1pt]
		(E6) & $P\equiv P'$ & $\Rightarrow$ & $\mem{P}\equiv\mem{P'}$ \\[1pt]
		(E7) & $X\equals X$ & $\equiv$ & $\textbf{0}$ \\[1pt]
 		(E8) & $X\equals Y$ & $\equiv$ & $Y\texttt{=}X$ \\[1pt]
		(E9) & $X\equals Y,P$ & $\equiv$ & $P[Y/X]$ &
                       if $P$ is an atom and $X$ is a free link of $P$ \\[1pt]
		(E10) & $\mem{X\equals Y,P}$ &
                $\equiv$ & $X\equals Y, \mem{P}$ & 
                 \hspace*{-14pt}
                 if exactly one of $X$ and $Y$ is a free link of $P$ \\[4pt]
		\hline
	\end{tabular}
	\caption{Structural congruence on LMNtal processes}
	\label{fig:Structural congruence on LMNtal processes}
\end{figure}

(E1)--(E3) characterize atoms as multisets.  
(E4) is  $\alpha$-conversion of local link names, where $Y$ must be a
fresh link because of the Link Condition.
(E5) and (E6) are
structural rules to make $\equiv$ a congruence.
(E7)--(E10) are rules for connectors.
\newsentence{They play key roles in many LMNtal programs, but in this paper}
we just refer the
readers to \cite{LMNtal} because connectors do not appear in this paper.


For 
convenience, the following term notation is introduced.
\begin{definition}[Term Notation]\label{def:termnotation}
We allow
$$
p(X_1,\narrowdots,X_{k-1},L,X_{k+1},\narrowdots,X_m ),\,q(Y_1,\narrowdots,Y_{n-\
1},L)\
(1 \leq k \leq m,1 \leq n),
$$
to be written as
$p(X_1,\narrowdots,X_{k-1},q(Y_1,\narrowdots,Y_{n-1}),X_{k+1},\narrowdots,X_m).$\qed
\end{definition}
For instance, 
\texttt{p(X,Y),a(X),b(Y)} 
can be written also as
\texttt{p(a,Y),b(Y)}
and then as
\texttt{p(a,b)},
which is much more concise and looks like a standard term.

\subsubsection{Reduction Relation}

\begin{figure}[tbp]
	\centering
	\begin{tabular}{ll}
		\hline\\[-2ex]
(R1) & $\dfrac{ P \reducesR{R} P'
  }{ P,Q \reducesR{R} P',Q
  }$\qquad\qquad\qquad
(R3) $\dfrac{ Q \equiv P
    \quad P \reducesR{R} P'
    \quad P' \equiv Q'
  }{ Q \reducesR{R} Q'
  }$\\[12pt]

(R4) & $\mem{X\equals Y,P}\reducesR{R} X\equals Y,\mem{P}$\quad if
                $X$ and $Y$ are free links of $\mem{X\equals Y,P}$ \\[3pt]
(R5) & $X\equals Y,\mem{P}\reducesR{R}
                \{X\equals Y,P\}$\quad if $X$ and $Y$ are free links of $P$ \\[3pt]

(R6) & $T\theta \reducesR{T \react U} U\theta$\\[3pt]
		\hline
	\end{tabular}
	\caption{Reduction relation on LMNtal processes}
	\label{fig:Reduction relation on LMNtal processes}
\end{figure}

The reduction relation by a rule $R$, denoted $\reducesR{R}$,
is the minimum binary relation satisfying the rules
in Fig.~\ref{fig:Reduction relation on LMNtal processes}.

(R1) and (R3) are structural rules.%
\footnote{\newsentence{(R2) for handling autonomous process evolution
  inside a membrane is irrelevant in the present formulation (which
  handles global rewrite rules only) and is omitted.  In the present
  setting, rewriting of the contents of a membrane must be specified
  as global rewrite rules that explicitly mention the membrane.}}
(R4) and (R5) handle interaction between connectors and membranes, and
we leave the details to \cite{LMNtal}.
The key rule, (R6), handles rewriting by a rule.  The substitution
$\theta$ handles how the process contexts in $T$ are instantiated;
it is a finite set $\{P_1/\texttt{\$}p_1,P_2/\texttt{\$}p_2,\dots\}$
of substitution elements of the form
$P_i/\texttt{\$}p_i$
that assigns $P_i$ to each process
context $\texttt{\$}p_i$ occurring in $T$.
Finally, given a set $S$ of rewrite rules, we write 
$P \reducesR{S}
P'$ if $\exists R\in S, P\reducesR{R} P'$.
\newsentence{%
We also write
$P \hspace{14pt}{\not}\hspace{-14pt}\reducesR{\raise2pt\hbox{\scriptsize
    $T\!\react\! U$}}$
if $\lnot\exists P'(P \reducesR{T\react U}P')$.
Since the body $U$ of the rule is irrelevant to
this non-reducibility property, we may also write it as
$P \hspace{10pt}{\not}\hspace{-10pt}\reducesR{\raise2pt\hbox{\scriptsize
$T\!\react$}}$, meaning that $P$ is not reducible by a rule with the
head $T$.}

%% file: tex/sec3.tex
\section{Introductory Examples of QLMNtal}
\label{sec:introductoryexamples}

Before formally defining the syntax and the semantics of QLMNtal, we
introduce its features by means of simple examples.  Further examples
will be given in Section~\ref{sec:more_examples},
and detailed explanation of how the examples work is given in 
\newsentence{a full version of this paper at arXiv.org}
(Appendix~\ref{sec:appendix_explanation}).

\subsection{Cardinality Quantification}
\label{subsec:Existential Quantification}

In QLMNtal, cardinality quantification tells how many
copies of the specified process may appear within the process to be
rewritten, and the matching is performed in a non-greedy manner
in the sense that
a cardinality-quantified process in the head
can match (or `grab')
\textit{any} number (within the specified range) of processes.
Then 
the corresponding cardinality-quantified process in the
body, if any, represents the same number of processes as the number of
processes grabbed by the cardinality-quantified process in the head.
For example, the QLMNtal rule
$$\texttt{<1,3>a(X,Y) :- <1,3>c(X,Y)}$$
says ``rewrite one to three copies of (binary) $\texttt{a}$ to
(binary) $\texttt{c}$.''

\begin{figure}[t]
	\begin{minipage}[b]{0.48\linewidth}
		\begin{screen}[4]
			\begin{minipage}[b]{0.6\linewidth}
				\begin{verbatim}
a(V,W),b(W,X),
a(X,Y),a(Y,Z),
b(Z,V)
				\end{verbatim}
			\end{minipage}
			\begin{minipage}[b]{0.35\linewidth}
				\centering
				\includegraphics[height=1.2cm]{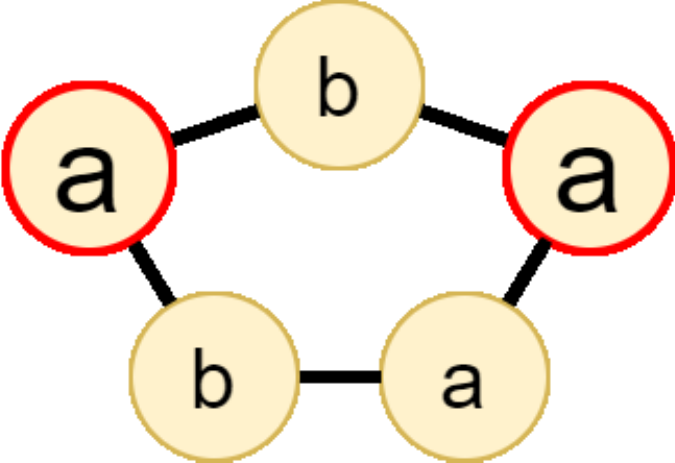}
				\vspace{-0.1cm}
			\end{minipage}
		\end{screen}
		\subcaption{Process before rewriting}
	\end{minipage}
\hfill
	\begin{minipage}[b]{0.48\linewidth}
		\begin{screen}[4]
			\begin{minipage}[b]{0.6\linewidth}
				\begin{verbatim}
c(V,W),b(W,X),
c(X,Y),a(Y,Z),
b(Z,V)
				\end{verbatim}
			\end{minipage}
			\begin{minipage}[b]{0.35\linewidth}
				\centering
				\includegraphics[height=1.2cm]{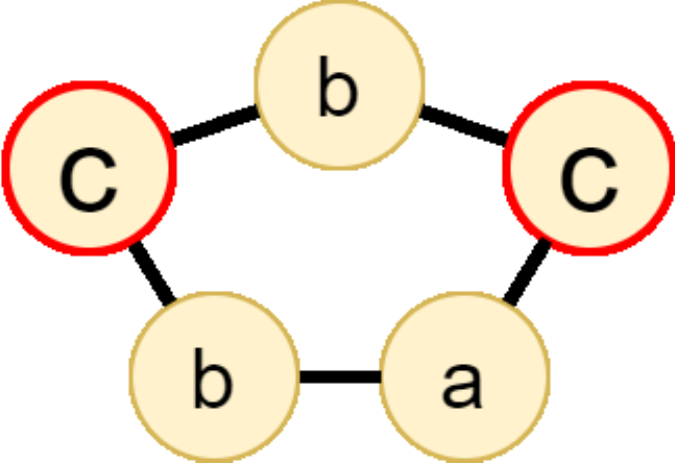}
				\vspace{-0.1cm}
			\end{minipage}
		\end{screen}
		\subcaption{Process after rewriting}
	\end{minipage}
	\caption{Rewriting with the rule \texttt{<1,3>a(X,Y) :- <1,3>c(X,Y)}}
	\label{fig:QLMNtal Program using Existential Quantification}
        \vspace*{-8pt}
\end{figure}

Two notes are appropriate here:
\begin{enumerate}
  \item The semantics in Section~\ref{sec:semantics}
        appropriately renames links of the three copies.
  \item The two quantifiers are 
interdependent;
as will be explained
    in Section~\ref{sec:semantics}, they are given the same `empty'
    label indicating the dependence between quantifiers.
\end{enumerate}
%
Figure~\ref{fig:QLMNtal Program using Existential Quantification}
shows an example of rewriting two $\texttt{a}$'s to two $\texttt{c}$'s.
Since the rewriting is nondeterministic,
it is also possible that one or three $\texttt{a}$'s are rewritten.

A cardinality quantifier $\texttt{<1,3>}$
specifies the range of the number of matching processes.
When the range is a single number, one may write $\texttt{<2>}$ instead of
$\texttt{<2,2>}$.
%
%
When one allows any cardinality, the quantifier can be written
as $\texttt{<?>}$.

\subsection{Non-existence Quantification}
\label{Non-existenceQuantification}

\begin{figure}[t]
	\begin{minipage}[b]{0.48\linewidth}
		\begin{screen}[4]
			\begin{minipage}[b]{0.5\linewidth}
				\begin{verbatim}
{a,b,b},
{a,b},
{a}
				\end{verbatim}
			\end{minipage}
			\begin{minipage}[b]{0.45\linewidth}
				\centering
				\includegraphics[height=1.2cm]{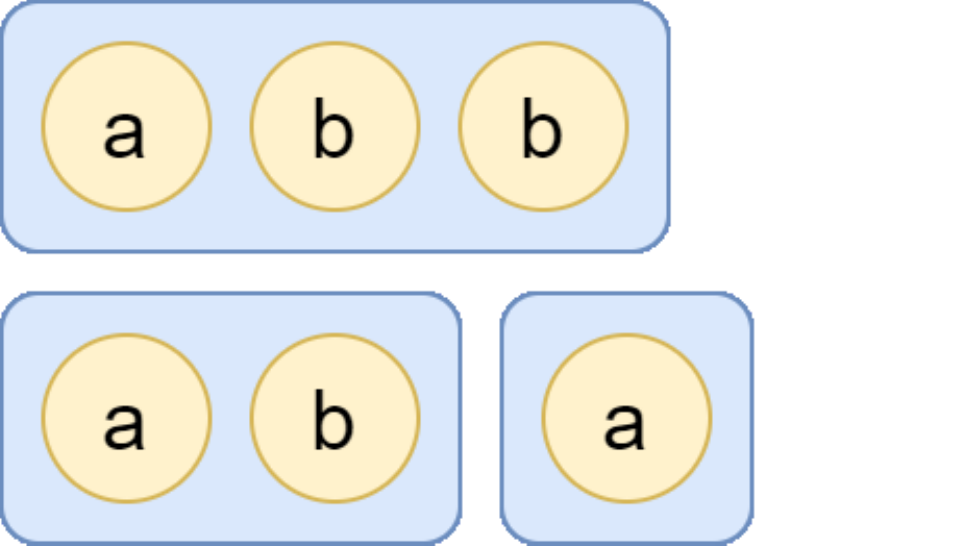}
				\vspace{-0.1cm}
			\end{minipage}
		\end{screen}
		\subcaption{Process before rewriting}
	\end{minipage}
\hfill
	\begin{minipage}[b]{0.48\linewidth}
		\begin{screen}[4]
			\begin{minipage}[b]{0.5\linewidth}
				\begin{verbatim}
{a,b,b},
{a,b},
{a},ok
				\end{verbatim}
			\end{minipage}
			\begin{minipage}[b]{0.45\linewidth}
				\centering
				\includegraphics[height=1.2cm]{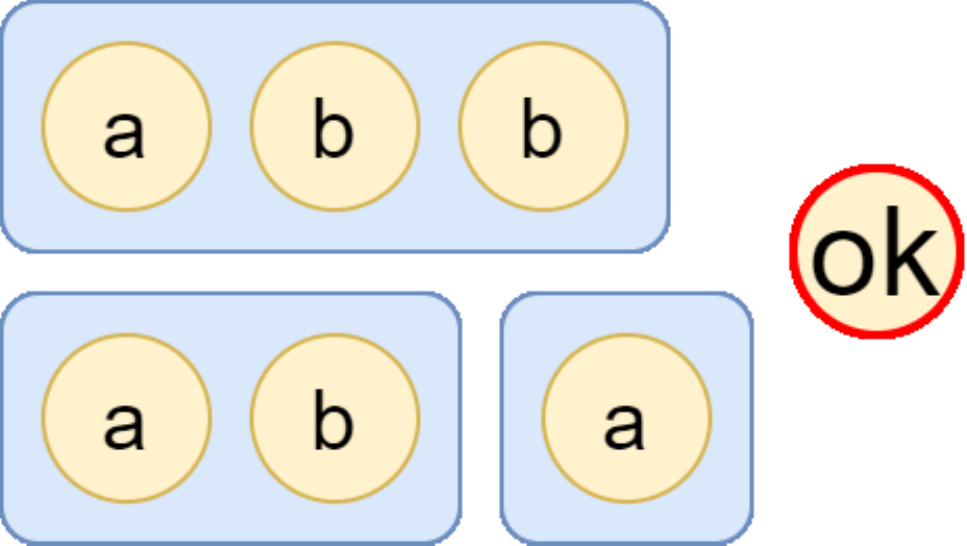}
				\vspace{-0.1cm}
			\end{minipage}
		\end{screen}
		\subcaption{Process after rewriting}
	\end{minipage}
	\caption{Rewriting with the rule \texttt{<\textasciicircum>\{a,a,\$p\} :- ok}}
	\label{fig:QLMNtal Program using Non-existence Quantification}
        \vspace*{-10pt}
\end{figure}

In QLMNtal, non-existence quantification is the basic construct for
expressing so-called negative application conditions (NACs).
For example, the rule
$$\verb|<^>{a,a,$p} :- ok|$$
%
\mishinasentence{says ``if no membrane contains two 
or more $\texttt{a}$'s, generate an
$\texttt{ok}$,''}
where $\texttt{<\textasciicircum>}$ stands for a non-existence quantifier.
Since non-existence quantification is 
\newsentence{used}
to express matching conditions,
it is meaningless to use non-existence quantification explicitly in
the body, and 
the Link Condition in Section~\ref{sec:syntax_LMNtal} does not apply
to 
links occuring under non-existence quantification.
\mishinasentence{
Figure~\ref{fig:QLMNtal Program using Non-existence Quantification}
shows an example of generating an $\texttt{ok}$ because a membrane
containing two or more $\texttt{a}$'s does not exist. 
}

\subsection{Universal Quantification}\label{UniversalQuantification}

Universal quantification in QLMNtal implies greedy matching,
that is,
a universally quantified process in the head
grabs
the largest possible number of processes,  
and the corresponding universally-quantified process in the body
represents the 
same number of processes as the number of processes 
grabbed 
in the head.
For example, the rule
$$\verb|<*>a(X,Y) :- <*>c(X,Y)|$$
says ``rewrite all binary $\texttt{a}$'s to binary $\texttt{c}$'s.''
This rule rewrites
$$\verb|a(V,W),b(W,X),a(X,Y),a(Y,Z),b(Z,V)|\quad
\textrm{(Fig.~\ref{fig:QLMNtal Program using Existential Quantification}(a))}$$
to
$\verb|c(V,W),b(W,X),c(X,Y),c(Y,Z),b(Z,V)|\ .$

It is important to note that the quantifier $\texttt{<*>}$ is provided
for notational 
convenience in the sense that the same rule can be written also as
$$\verb|<?>a(X,Y),<^>a(X0,Y0) :- <?>c(X,Y),<^>c(X0,Y0)|\ , $$
where the non-existence quantification tells that the 
$\verb|<?>a(X,Y)|$ should grab the largest possible number of the binary
$\texttt{a}$'s.
That is, if
\verb|<?>a(X,Y)|
does not grab all the $\texttt{a}$'s,
the pattern matching fails due to $\verb|<^>a(X0,Y0)|$.

QLMNtal provides another (derived) universal quantifier, $\texttt{<+>}$,
corresponding to $\forall^{>0}$ in logic, which is useful when one
wishes to avoid vacuous rule application when there are zero instances
of the specified process.

\subsection{Combined Use of Quantifiers}\label{CombinedUse}

QLMNtal allows quantifiers to be used in combination.
For example, the rule
$$\verb|<*>{a,b,$p1},<^>{b,$p2} :- <*>{a,b,$p1},ok|$$
%
says ``generate an \texttt{ok} if all membranes containing
   \texttt{b} contain \texttt{a} as well.''
This rule rewrites 
\verb|{a,b,b},{a,b},{a}|
(Fig.~\ref{fig:QLMNtal Program using Non-existence Quantification}(a))
to
$\verb|{a,b,b},{a,b},{a},ok| .$

%% file: tex/sec4.tex
\section{Syntax of QLMNtal}
\label{sec:syntax}

We first introduce the following syntactic convention.

\begin{definition}[vector notation]
For some syntactic entity $E$, $\overrightarrow{E}$ stands for a
sequence $E_1,\dots ,E_n$ for some $n\ (\geq 0)$.  
When we wish to mention the indices explicitly, $E_1,\dots,E_n$ will
also be denoted as ${\overrightarrow{E_i}}^i$.\qed
\end{definition}


The syntax of QLMNtal is shown in Fig.~\ref{fig:Syntax of QLMNtal},
which adds 
to the syntax of LMNtal (Fig.~\ref{fig:Syntax of LMNtal}) 
notations for the cardinality quantifier and the non-existence
quantifier for process templates.  Note that the syntax of processes
remains exactly the same and is omitted.
In the syntax of the quantifier $\Q$, 
\begin{itemize}
\item $l$ stands for a \textit{quantifier
  label} whose concrete syntax is either (i) an empty name or (ii) a name
  starting with a capital letter;
\item $z$ stands for an integer or
  $\infty$, where we assume $\infty-1=\infty$
  and $\forall n\,(n<\infty)$;
\item $\Q$ binds tighter than `$,$' for parallel
  composition; for example, $\texttt{<3,5>a,a,a}$ stands for
  $\texttt{(<3,5>a),a,a}$ and not $\texttt{<3,5>(a,a,a)}$.
\end{itemize}

\begin{figure}[t]
	\centering
	\begin{tabular}{rp{10pt}cll}
		\hline\\[-6pt]
(rule)          & $R$ & $::=$ & $T\react T$ & \\[3pt]
(template)	& $T$ & $::=$ & $\mathbf{0}$ & (null) \\
		& & $|$ & $p(X_1,\dots,X_m)$\quad$(m\ge 0)$ & (atom) \\
		\rowcolor{mygray}
		& & $|$ & $\Q T$ & (quantified template) \\
		& & $|$ & $T,T$ & (parallel composition) \\
		& & $|$ & $\mema{T}$ & (membrane) \\
		& & $|$ & $\texttt{\$}p$ & (process context) \\[3pt]
		\rowcolor{mygray}
(quantifier)	& $\Q$ & $::=$ & $\carda{l}{z}{z}{}$ & (cardinality quantifier) \\
		\rowcolor{mygray}
		& & $|$ & \newsentence{$\numa{l}{\texttt{\textasciicircum}}{}$} & (non-existence quantifier) \\[3pt]
		\hline
	\end{tabular}
	\caption{Syntax of QLMNtal}
	\label{fig:Syntax of QLMNtal}
\end{figure}


\noindent
$\Q T$ is a construct encompassing the 
\textit{cardinality quantification template}
$\carda{l}{z_1}{z_2}{T}$  
and the \textit{non-existence quantification template}
$\numa{l}{\texttt{\textasciicircum}}{T}$.
In a quantified template $\Q T$, $\Q$ is called the \textit{(labeled)
quantifier} and 
$T$ is called the \textit{quantified part}.  
Quantified templates can be nested as in
$\Q_0 \Q_1 \Q_2 T$.
The quantifier of a quantified template that 
\newsentence{is}
not inside another
quantifier is called the \textit{outermost quantifier}.
In the sequel, $\Q_0$ (a quantifier with the suffix 0)
denotes an outermost quantifier.

A quantified template must observe the following syntactic conditions.

\begin{definition}[Syntactic Conditions for Quantified Templates]
Links or 
process contexts appearing in a quantified template 
\newsentence{may}
appear only
in (the same or different) quantified templates with the same
quantifier.\qed
\end{definition}

An important feature of QLMNtal is the \textit{labeling of
quantifiers}.  We often wish to \textit{jointly} quantify entities
distributed over isolated places of the membrane structure and
entities on both sides (head and body) of a rule.  However, it is
difficult to achieve this in textual syntax.  Accordingly, we introduce
\textit{labels} to associate physically distributed but logically
related quantifiers.
For instance, in \verb|L<^>a(X),{L<^>b(X),$p}|, $\texttt{a(X)}$ and
$\texttt{b(X)}$ are quantified collectively.
%

\subsubsection{Cardinality Quantification Template}

The cardinality quantification template $\carda{l}{z_1}{z_2}{T}$ 
stands for either of $z_1,\ldots,z_2$ copies of $T$
by the extended structural congruence rule to be given in Section~\ref{sec:SC},
where the links 
of the $T$'s are appropriately renamed by (E4).
%
The $l$ is a (possibly empty) label name used to distinguish between
independent cardinality quantifiers.

%
Note that a cardinality quantification template does not itself
enforce a greedy match.
For example, given a process \texttt{a,a,a,a}, the template
\texttt{<1,3>a} can match either one, two, or three of the four
\texttt{a}'s.

Table~\ref{tab:Notations of Existential Quantification Template}
summarizes the notations of cardinality quantification templates.

\begin{table}[b]
        \vspace*{-10pt}
	\caption{Notations of Cardinality Quantification Templates}
	\centering
        \vspace*{3pt}
	\begin{tblr}{@{\,}l@{\,}|@{\,}l@{\,}|@{\,}l@{\,}|@{\,}l@{\,}}
	\hline
	Notation & Shorthand & Meaning & Example \\ \hline[1pt]
	$\carda{l}{0}{\infty}{T}$ & $\numa{l}{\texttt{?}}{T}$ & any number of $T$'s & \texttt{<?>a(X)} \\ \hline
	$\carda{l}{z}{\infty}{T}$ & $\carda{l}{z}{}{T}$ & not less than
        $z$ $T$'s & \texttt{<3,>a(X)} \\ \hline
	$\carda{l}{z_1}{z_2}{T}$ & $\carda{l}{z_1}{z_2}{T}$ & not less
        than $z_1$ and not more than $z_2$ $T$'s & \texttt{<3,5>a(X)} \\ \hline
	$\carda{l}{z}{z}{T}$ & $\numa{l}{z}{T}$ & exactly $z$ $T$'s &
        \texttt{<3>a(X)} \\ \hline
	\end{tblr}
	\label{tab:Notations of Existential Quantification Template}
\end{table}

The outermost cardinality quantifier, 
denoted as $\carda{l}{z_1}{z_2}{}_0$,
plays an important role in the structural congruence and the reduction
relation
described in Section~\ref{sec:semantics}.

\subsubsection{Non-existence Quantification Template}

$\numa{l}{\texttt{\textasciicircum}}{T}$ is a template that indicates the
non-existence of a process,
and the pattern matching fails if there is a process that matches the
template $T$.
The $l$ is a (possibly empty) label name used to distinguish between
unrelated non-existence quantifiers.
(Non-existence and cardinality quantifiers are never
related to each other.)
The outermost non-existence quantifier,
denoted as $\numa{l}{\texttt{\textasciicircum}}{}_0$, plays an important role in
the reduction relation.

We note that
$\numa{l}{\texttt{\textasciicircum}}{T}$ appearing in the body
may be omitted \newsentence{as already explained in
Section~\ref{Non-existenceQuantification}.}
For example, 
$\texttt{<\textasciicircum>a(X)}\mathop{\texttt{:-}}\texttt{<\textasciicircum>b(X)}$
can be written also as 
$\texttt{<\textasciicircum>a(X)}\mathop{\texttt{:-}}\texttt{.}$
This is because the non-existence quantification in the body of a rule 
does not play any role (other than to satisfy the occurrence
conditions of free links and process contexts) in the semantics
defined in Section~\ref{sec:semantics}.
This abbreviation may omit free links and process
contexts in the body, but we do not consider it as violating
the syntactic conditions of rules.

\subsection{Representation of Universal Quantification}

As explained in Section~\ref{UniversalQuantification},
universal quantification of QLMNtal
can be represented
by combining cardinality quantification and non-existence
quantification.
%
%
To enable concise description,
we provide shorthands $\numa{}{\texttt{*}}{}$ and $\numa{}{\texttt{+}}{}$
given in Table~\ref{tab:Notations for expressing
  Universal Quantification}.
In order to satisfy the Link Condition and the
occurrence conditions of process contexts (Section~\ref{sec:syntax_LMNtal}),
$T'$ stands for a process template obtained from $T$ by renaming
links and process contexts using fresh names.
%
Likewise,
each label name $l$ is renamed to its `primed' version $l'$ to distinguish it 
from the label name of the cardinality quantification template and the
non-existence quantification template. 
This way, if the same label name $l$ is used to write
$\numa{l}{\texttt{*}}{T_1},\numa{l}{\texttt{\textasciicircum}}{T_2}$
\newsentence{as in the example of Section~\ref{CombinedUse}},
the label names do not conflict 
because they are translated to $(\carda{l'}{0}{\infty}{T_1},
\numa{l'}{\texttt{\textasciicircum}}{T_1},\numa{l}{\texttt{\textasciicircum}}{T_2}$).

\begin{table}[b]
        \vspace*{-10pt}
	\caption{Notations for expressing Universal Quantification}
	\centering
        \vspace*{3pt}
        \begin{tblr}{l@{~}|@{~}l@{~}|@{~}l@{~}|@{~}l}
	\hline
	Notation & Shorthand & Meaning & Example \\ \hline[1pt]
        $\carda{l'}{0}{\infty}{T},
         \newsentence{\numa{l'}{\texttt{\textasciicircum}}{T'}}$ &
        $\numa{l}{\texttt{*}}{T}$ & maximum number of $T$'s & \texttt{<*>a(X)} \\ \hline
        $\carda{l'}{1}{\infty}{T},
         \newsentence{\numa{l'}{\texttt{\textasciicircum}}{T'}}$ &
        $\numa{l}{\texttt{+}}{T}$ & maximum number ($\ge 1$) of $T$'s &
        \texttt{<+>a(X)} \\ 
	\hline
	\end{tblr}
	\label{tab:Notations for expressing Universal Quantification}
\end{table}

%% file: tex/sec5.tex
\section{Semantics of QLMNtal}
\label{sec:semantics}

\subsection{Structural Congruence of QLMNtal}
\label{sec:SC}

One of the key ideas towards using quantified rewrite rules is to give
a law for ``unrolling'' cardinality quantification.  This is achieved by
an additional equivalence relation (EQ) shown below:

\medskip\noindent
\fbox{%
	\begin{tabular}{l}
	(EQ)  $T\react U \ \equiv\
              (T\react U)\bigl[{\overrightarrow{%
              (\carda{l}{z_1\!-\!1}{z_2\!-\!1}{_0\,T_i},\,T'_i)\,\big/\,
              \carda{l}{z_1}{z_2}{_0\,T_i}}}^i\bigr]$\\[2pt]
where ${\strut\overrightarrow{\carda{l}{z_1}{z_2}{_0\,T_i}}}^i$ 
  stands for all the 
templates quantified with $\carda{l}{z_1}{z_2}{_0}$,\\[-2pt]
${\overrightarrow{T'_i}}^i$ stands for
templates in which all the links, process contexts,
and labels \\[-2pt]
appearing in ${\overrightarrow{T_i}}^i$ are replaced
  with fresh names not appearing in $T\react U$ \\
	\end{tabular}
}%

\medskip\noindent
(EQ) states that all $T_i$'s quantified by 
$\carda{l}{z_1}{z_2}{_0}$ 
\newsentence{(i.e., the outermost occurrences of the quantifier
$\carda{l}{z_1}{z_2}{}$)}
in a rule 
can be expanded to 
$\bigl(\carda{l}{(z_1\!-\!1)}{(z_2\!-\!1)}{_0\,T_i},\,T'_i\bigr)$.
That is, the outermost cardinality quantification template
$\carda{l}{z_1}{z_2}{_0\,T_i}$ can be unrolled by replicating
$T_i$ and decrementing $z_1$ and $z_2$.
In order to maintain consistency, 
\newsentence{all the outermost occurrences of the
cardinality
quantifier with the same $l,z_1,z_2$}
must be replicated at the same time. 
Furthermore, all the link names, process context names and label names
inside the replicated $T_i$ must be renamed with fresh names.

For example, the following rules are structurally congruent, where
the underline indicates the process template replicated by (EQ).

\medskip\noindent
\begin{tabular}{cl}
& \texttt{\card{}{0}{$\infty$}{\{a(X),\card{}{3}{3}{b},\$c\}}
  :- \card{}{0}{$\infty$}{\{a(X),\$c\}}}\\[-2pt]
$\stackrel{\textrm{(EQ)}}{\equiv}$ &
\texttt{\card{}{-1}{$\infty$}{\{a(X),\card{}{3}{3}{b},\$c\}},
\underline{\{a(X1),\card{M}{3}{3}{b},\$c1\}}}\\
& \texttt{:- \card{}{-1}{$\infty$}{\{a(X),\$c\}},\underline{\{a(X1),\$c1\}}}\\[-2pt]
$\stackrel{\textrm{(EQ)}}{\equiv}$ &
\texttt{\card{}{-1}{$\infty$}{\{a(X),\card{}{3}{3}{b},\$c\}},
\{a(X1),\card{M}{0}{0}{b},\underline{b,b,b},\$c1\}}\\
& \texttt{:- \card{}{-1}{$\infty$}{\{a(X),\$c\}},\{a(X1),\$c1\}}
\end{tabular}
\medskip

\noindent
Note that (EQ) itself
allows 
$\carda{l}{z_1}{z_2}{_0 T_i}$
to be unrolled more than
$z_1$ times, resulting in the `\texttt{-1}' in the above example.
The right number of unrolling is ensured by (\textit{CardCond}) shown in 
Section~\ref{subsubsec:CardCond}. 

\subsection{Reduction Relation of QLMNtal}

The reduction relation of QLMNtal is similar to that of LMNtal
(Fig.~\ref{fig:Reduction relation on LMNtal processes}) but with
changes to handle cardinality and non-existence.
While the reduction of LMNtal processes proceeds solely based on the
\textit{existence} of subprocesses that can be checked locally,
checking of \textit{non-existence} requires access to the entire
process.  In the terminology of programming language theory,
this means we
must (partly) switch from inductive
small-step semantics with structural rules 
(w.r.t. parallel composition)
to so-called \textit{contextual semantics}.  In mainstream
programming languages, contextual semantics is mainly used
for specifying the order of evaluation, but here we use it to access
global information.

Specifically,
in QLMNtal, the structural rule (R1) for parallel composition and
the main rule (R6) are deleted from the rules of
Fig.~\ref{fig:Reduction relation on LMNtal processes}. 
They are replaced with the new rule (RQ) and the accompanying
structural rule (R3$'$) below:


\medskip\noindent
\fbox{%
\begin{tabular}{lc}
	(RQ) & $\dfrac{%
  \forall \carda{l}{z_1}{z_2}{_0}\bigl(z_1\le 0\land z_2\ge 0\bigr)\hfill
  \ \land\; \forall \numa{l}{\texttt{\caret}}{}_0
  \bigl(\context(\mema{T,\texttt{\$}\self})\theta
  \hspace{30pt}{\not}\hspace{-30pt}
  \reducesR{
\raise3pt\hbox{\scriptsize$\negation(l,\mema{T,\texttt{\$}\self})
                      \!\react
$}}
%
\bigr)} 
%
{%
(\func(T),\pc\self)\theta \reducesR{T\react U} 
 (\func(U),\pc\self)\theta}$ \\[15pt] 
(R3$'$) & $\dfrac{R\equiv R'\quad P\reducesR{R}P'}{P\reducesR{R'}P'}$\\
	\end{tabular}
}

\medskip\noindent
(RQ) consists of (i) a conclusion representing the rewriting 
and (ii) a premise under which the rewriting takes place.
The $\pc\self$, called a \textit{\mishinasentence{global 
context}}, is a special process context that represents
the ``context'' of the head of the rule,
i.e., graph elements not affected by rewriting.
%
We could think of the existence of an \textit{(implicit) global membrane}
which contains all processes and in which all rewritings take
place; $\pc\self$ could be thought of the process context of
the global membrane.  Now the role of the substitution $\theta$ 
in (RQ) is extended to
instantiate $\pc\self$ as well as the other process contexts in the
head of the rule.

We will now explain how (RQ) works step-by-step.

\subsubsection{Rewriting of a Process}

Let (RQ$'$) be the conclusion of (RQ), i.e.,
%
%
$$(\func(T),\pc\self)\theta \reducesR{T\react U} 
 (\func(U),\pc\self)\theta\ .$$
%
%
%
The auxiliary function $\textit{simp}$ removes quantification
templates from the head and the body of a rule, 
whose definition is given in Fig.~\ref{fig:simplify()}.
%

\begin{figure}[t]
\centering
\begin{tabular}{rcl@{\quad}rcl}
\hline\\[-2ex]
$\func(\textbf{0})$ & $\stackrel{\mathrm{def}}{=}$ & $\textbf{0}$ &
$\func(p(\overrightarrow{X}))$ & $\stackrel{\mathrm{def}}{=}$ & $p(\overrightarrow{X})$ \\
$\func(\mema{T})$ & $\stackrel{\mathrm{def}}{=}$ & $\mema{\func(T)}$ &
$\func(\Q T)$ & $\stackrel{\mathrm{def}}{=}$ & \textbf{0} \\
$\func((T_1,T_2))$ & $\stackrel{\mathrm{def}}{=}$ & $(\func(T_1),\func(T_2))$ &
$\func(\pc p)$ & $\stackrel{\mathrm{def}}{=}$ & $\pc p$ \\[3pt]
		\hline
	\end{tabular}
	\caption{$\textit{simp}$ used in (RQ) and (RQ$'$)}
	\label{fig:simplify()}
\vspace*{-9pt}
\end{figure}

\begin{figure}[t]
\begin{minipage}[b]{0.47\linewidth}
	\begin{tabular}{rcl}
	\hline\\[-2ex]
	$\context(\textbf{0})$ & $\stackrel{\mathrm{def}}{=}$ & $\textbf{0}$ \\
	$\context(p(\overrightarrow{X}))$ & $\stackrel{\mathrm{def}}{=}$ & $\textbf{0}$ \\
	$\context(\mema{T})$ & $\stackrel{\mathrm{def}}{=}$ & $\mema{\context(T)}$ \\
	$\context(\Q T)$ & $\stackrel{\mathrm{def}}{=}$ & \textbf{0} \\
	$\context((T_1,T_2))$ & $\stackrel{\mathrm{def}}{=}$ & $(\context(T_1),\context(T_2))$ \\
	$\context(\pc p)$ & $\stackrel{\mathrm{def}}{=}$ & $\pc p,\itnameid{p}()$ \\[3pt]
	\hline
	\end{tabular}
	\caption{$\textit{cxt}$ used in (RQ)}
	\label{fig:context()}
\end{minipage}
\begin{minipage}[b]{0.51\linewidth}
	\begin{tabular}{rcl}
	\hline\\[-2ex]
	$\negation(l,\textbf{0})$ & $\stackrel{\mathrm{def}}{=}$ & $\textbf{0}$ \\
	$\negation(l,p(\overrightarrow{X}))$ & $\stackrel{\mathrm{def}}{=}$ & $\textbf{0}$ \\
	$\negation(l,\mema{T})$ & $\stackrel{\mathrm{def}}{=}$ & $\mema{\negation(l,T)}$ \\
	$\negation(l,\Q T)$ & $\stackrel{\mathrm{def}}{=}$ &
        $T$ \hspace{30pt}$(\Q=\numa{l}{\texttt{\caret}}{})$ \\
	$\negation(l,\Q T)$ & $\stackrel{\mathrm{def}}{=}$ & \textbf{0} \hspace{30pt}$(\Q\neq \numa{l}{\texttt{\caret}}{})$ \\
	$\negation(l,(T_1,T_2))$ & $\stackrel{\mathrm{def}}{=}$ & $(\negation(l,T_1),\negation(l,T_2))$ \\
	$\negation(l,\pc p)$ & $\stackrel{\mathrm{def}}{=}$ & $\pc\itnamecxt{p},\itnameid{p}()$ \\[3pt]
	\hline
	\end{tabular}
	\caption{$\textit{neg}$ used in (RQ)}
	\label{fig:negation()}
\end{minipage}
\vspace*{-8pt}
\end{figure}

For example, let $R = \texttt{\card{}{2}{2}{a}}\react
\texttt{\card{}{2}{2}{b}}$,
which says ``replacing two \texttt{a}'s with two \texttt{b}'s'', 
can be transformed to
$R'=\texttt{\card{}{0}{0}{a},a,a}\react\texttt{\card{}{0}{0}{b},b,b}$
by using (EQ) twice
and then be applied using (RQ$'$) as follows: 
%

$$\begin{tabular}{rl}
&\texttt{a,a,a}\
$\bigl(\equiv$ $(\func(\texttt{\card{}{0}{0}{a},a,a}),\pc\self)
 [\texttt{a}/\pc\self]\;\bigr)$\\
$\reducesR{R'}$&
\texttt{b,b,a}\
$\bigl(\equiv$ $(\func(\texttt{\card{}{0}{0}{b},b,b}),\pc\self)
 [\texttt{a}/\pc\self]\;\bigr)$.
\end{tabular}$$

\subsubsection{Condition on the outermost cardinality quantifier} 
\label{subsubsec:CardCond}

The condition $z_1\le 0\;\land\;z_2\ge 0$ in (RQ), hereafter referred
to as $(\textit{CardCond\/})$, is a condition that all 
outermost quantifiers
$\carda{l}{z_1}{z_2}{}$ (the suffix `$_0$' in (RQ) standing for `outermost')
in the rule must satisfy when applying the rule.
%
%
(\textit{CardCond}) confirms that the quantified template $T$ of
$\carda{l}{z_1}{z_2}{T}$ is replicated
between $z_1$ and $z_2$ times by (EQ).
For example, the rule $\texttt{\card{}{2}{2}{a}}\react\texttt{\card{}{2}{2}{b}}$
described above is equivalent to 
$\texttt{\card{}{1}{1}{a},a}\react\texttt{\card{}{1}{1}{b},b}$,
but this rule
cannot be used for rewriting because it does not satisfy
(\textit{CardCond}). 

\subsubsection{Condition on the outermost non-existence quantifier}
\label{subsubsec:NegCond}

The following (\textit{Neg\-Cond}), the second premise of (RQ),
states what all outermost non-existence quantifiers
$\numa{l}{\texttt{\caret}}{}$ 
in the rule must satisfy when used for
rewriting:

$$\textrm{(\textit{NegCond})}\ 
 \forall \numa{l}{\texttt{\caret}}{}_0
  \bigl(\context(\mema{T,\texttt{\$}\self})\theta
  \hspace{30pt}{\not}\hspace{-30pt}
  \reducesR{
\raise3pt\hbox{\scriptsize$\negation(l,\mema{T,\texttt{\$}\self})
                      \!\react
$}}
%
\bigr)\ ,
$$

%
\noindent
where $\context$ and $\negation$ are defined in Fig.~\ref{fig:context()}
and Fig.~\ref{fig:negation()}, respectively.

(\textit{NegCond}) confirms that there is no process matching the
non-existence quantifier in each of the ``places'' of the membrane
hierarchy.
%

Intuitively,
$\context(\mema{T,\pc\self})\theta$ represents
the `target' of the non-existence check, 
where the target means 
all elements whose existence is not explicitly
mentioned in the head (but is represented by process contexts
including $\pc\self$).
%
%
$\negation(l,\mema{T,\pc\self})$ extracts all
the quantified parts of $\numa{l}{\texttt{\caret}}{}$.
Pattern matching is performed between these two, 
$\context(\mema{T,\pc\self})\theta$ and
$\negation(l,\mema{T,\pc\self})$,
and its success/failure
indicates that the non-existence condition 
prefixed with
$\numa{l}{\texttt{\caret}}{}$ is unsatisfied/satisfied, respectively.


The function $\context$ (Fig.~\ref{fig:context()})
removes all atoms and quantified templates 
from the process
template while maintaining the original membrane hierarchy.
Thus, $\context(\mema{T,\pc\self})$
represents the extracted process contexts in the hierarchy.
By instantiating those process contexts by $\theta$,
the context of the head of the rule will be extracted.
It also generates a unique \textit{id atom} in each membrane containing
a process context.
The purpose of the id atoms is to identify each child membrane when
there are 
multiple child membranes within the same parent membrane. 

The function $\negation$ (Fig.~\ref{fig:negation()})
removes all atoms and cardinality quantifications
from the process template while maintaining the original membrane
hierarchy, and removes the top-level non-existence quantifiers with
the specified label $l$.
Thus, $\negation(l, \mema{T,\pc\self})$
represents the elements that should \textit{not} exist in each
place of the membrane structure.
In addition,
$\negation$ renames process contexts with fresh names
(denoted as $\pc p^{ct}$ in Fig.~\ref{fig:negation()})
(to avoid name crash with existing ones) and generates a
unique id atom in each membrane containing a process context, e.g.,
$$	\negation(\texttt{M},\texttt{\{\{M\caret a(X),\$p\},M\caret a(X),\$$\self$\}})
	=\texttt{\{\{a(X),\$\ttnamecxt{p},\ttnameid{p}\},a(X),\$\ttnamecxt{$\self$},\ttnameid{$\self$}\}}\ .$$
For example, let
$$R = \texttt{\{M<\caret>a(X),\$p\},}\ \texttt{M<\caret>a(X)}
 \react\texttt{\{\$p\},}\ \texttt{ok}, $$
stating ``generate an \texttt{ok} when there are no \texttt{a}'s
connected across any membrane.'' If (RQ$'$) is applied to $R$,
$R$ could perform the following rewriting,
where
$\theta=\bigl[(\texttt{a(X),b})/\texttt{\$p},\,
              (\texttt{a(X),c})/\pc\self\bigr]$:


\medskip\noindent
\hspace{19pt}%
\texttt{\{a(X),b\},a(X),c.}\
\hspace{12pt}
($\equiv (\func(\texttt{\{M<\caret>a(X),\$p\},M<\caret>a(X)}),\pc\self)
\theta$\;)\\
%
$\reducesR{R}$ \texttt{\{a(X),b\},a(X),c,ok.}\
($\equiv (\func(\texttt{\{\$p\},ok}),\pc\self)
\theta$\;)\\
\smallskip

\noindent
However, no such rewriting is performed by (RQ) because
the outermost non-existence quantifier \texttt{M<\caret>} does
not satisfy (\textit{NegCond}):

\medskip\noindent
(\textit{NegCond}) for the \texttt{M<\caret>}\\[1pt]
$= \context(\texttt{\{\{M<\caret>a(X),\$p\},M<\caret>a(X),\$$\self$\}})\theta
\hspace{60pt}{\not}\hspace{-60pt}
\reducesR{\raise4pt\hbox{\scriptsize$
\negation\bigl(\texttt{M},\texttt{\{\{M<\caret>a(X),\$p\},M<\caret>a(X),\$$\self$\}}\bigr)\react
$}}
$\\[2pt]
$= \texttt{\{\{\underline{a(X)},b,\ttnameid{p}\},\underline{a(X)},c,\ttnameid{$\self$}\}}
\hspace{65pt}{\not}\hspace{-65pt}
\reducesR{\raise4pt\hbox{\scriptsize$
\texttt{\{\{a(X),\$\ttnamecxt{p},\ttnameid{p}\},a(X),\$\ttnamecxt{$\self$},\ttnameid{$\self$}\}}\react
$}}$\\[4pt]
$=\textit{false}\ .$
%
%

\medskip
Notice that the definition of (\textit{NegCond}) is inductive because
it refers to the reduction relation of QLMNtal.
This means that we only need to explicitly check that the outermost
non-existence quantifier satisfies 
(\textit{NegCond}); the inner quantifiers will be checked by recursion.
Note also that the inductive definition of (\textit{NegCond}) is
well-founded because each application of $\negation$ removes the
non-existence quantifier by exactly one level.

%% file: tex/sec6.tex
\section{More Examples}
\label{sec:more_examples}

\subsection{Repotting the Geraniums}
\label{subsec:Repotting the Geraniums}

\begin{figure}[ht]
        \vspace*{-12pt}
	\begin{minipage}[b]{0.49\linewidth}
	\centering
	\includegraphics[width=4cm]{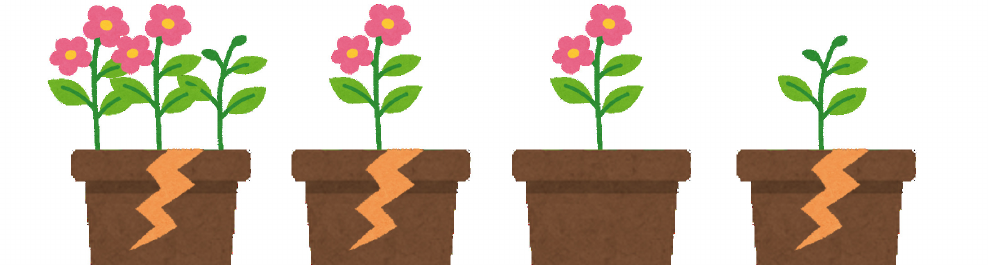}
	\subcaption{\newsentence{Before repotting}}
	\end{minipage}
	\begin{minipage}[b]{0.49\linewidth}
	\centering
	\includegraphics[width=4cm]{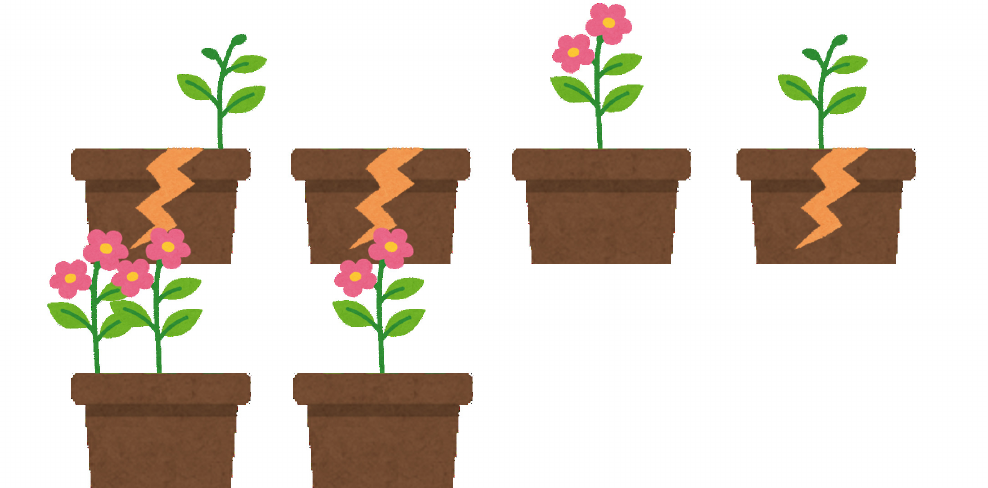}
	\subcaption{\newsentence{After repotting}}
	\end{minipage}
	\caption{\newsentence{Repotting the Geraniums}}
	\label{fig:Repotting the Geraniums}
\end{figure}

``Repotting the Geraniums'' \cite{Geraniums} is a 
problem
that requires a nested quantification rule, 
\newsentence{%
whose description is given below and visualized in 
Fig.~\ref{fig:Repotting the Geraniums}:}

\begin{quote}\textsl{%
    There are several pots, each with several geranium plants.
    Some 
    pots were broken because the geraniums filled the 
    space with their roots.
    New pots are prepared for the broken pots with flowering geraniums
    and all the flowering
    geraniums are moved
    to the new pots.}\>
    ---~\cite{Geraniums} 
\end{quote}

\noindent
\newsentence{%
The original modeling of the problem in a visual graph rewriting tool
GROOVE \cite{GROOVE} is shown in 
Fig.~\ref{fig:Repotting the Geraniums in GROOVE}.}

An example of QLMNtal modelling of this problem is 
shown below:
%

\begin{figure}[t]
	\begin{minipage}[b]{0.265\linewidth}
	\centering
	\includegraphics[width=3.2cm]{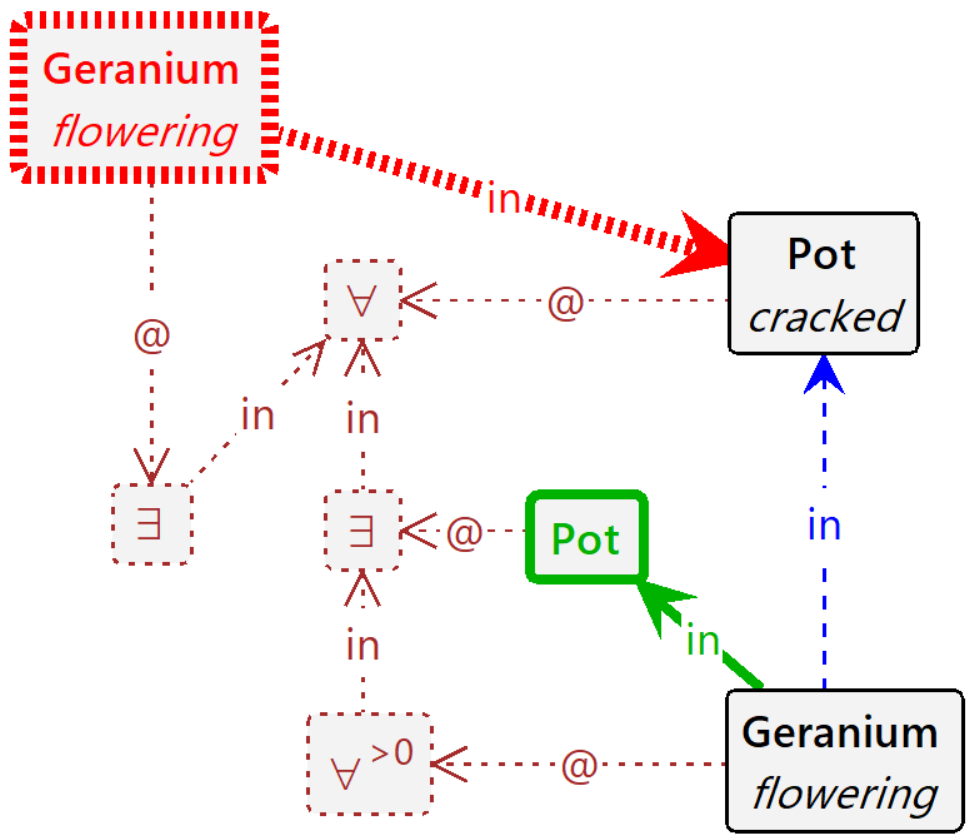}
	\subcaption{GROOVE rule}
	\end{minipage}%
	\begin{minipage}[b]{0.365\linewidth}
	\centering
	\includegraphics[width=4.45cm]{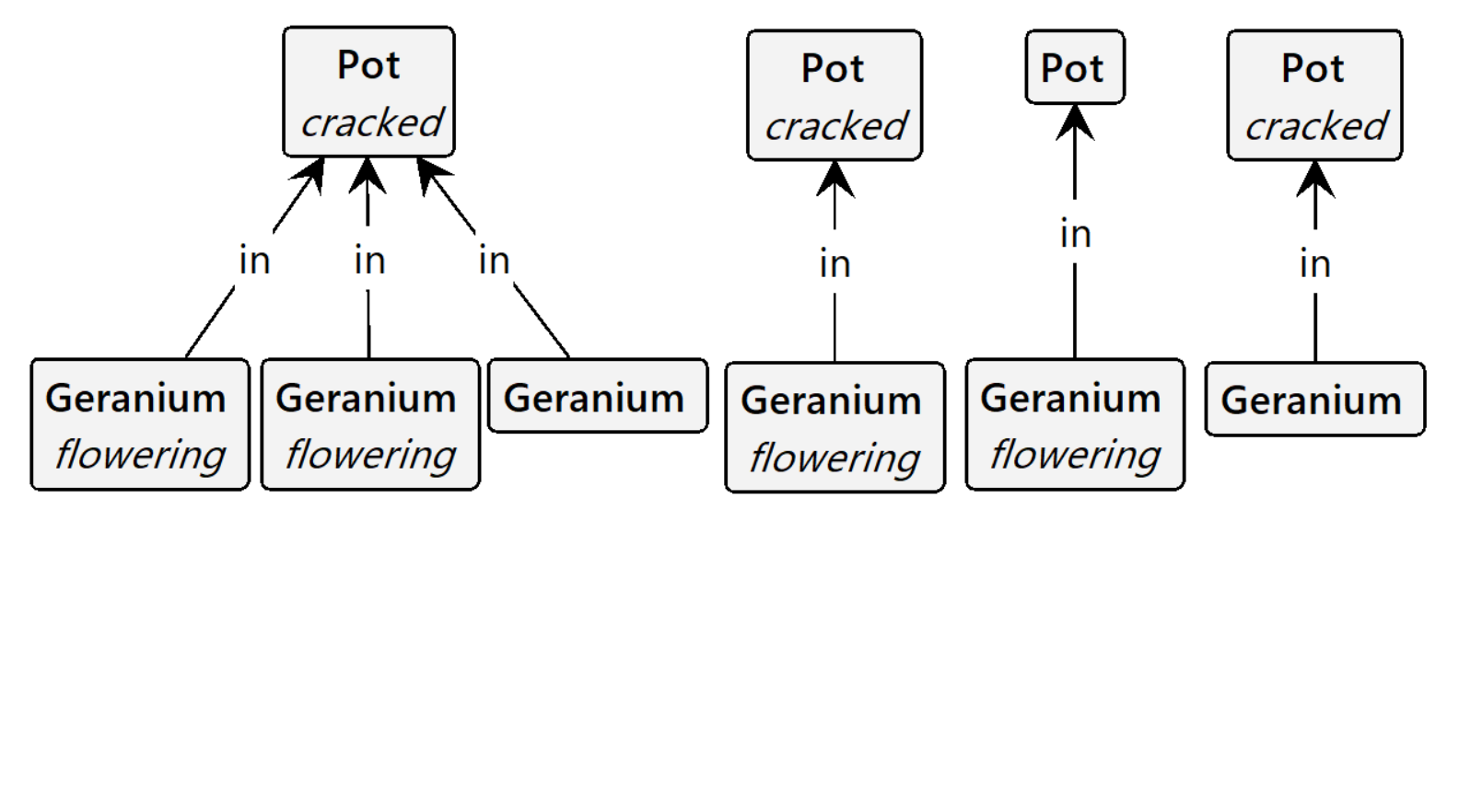}
	\subcaption{Graph before rewriting}
	\end{minipage}%
	\begin{minipage}[b]{0.365\linewidth}
	\centering
	\includegraphics[width=4.45cm]{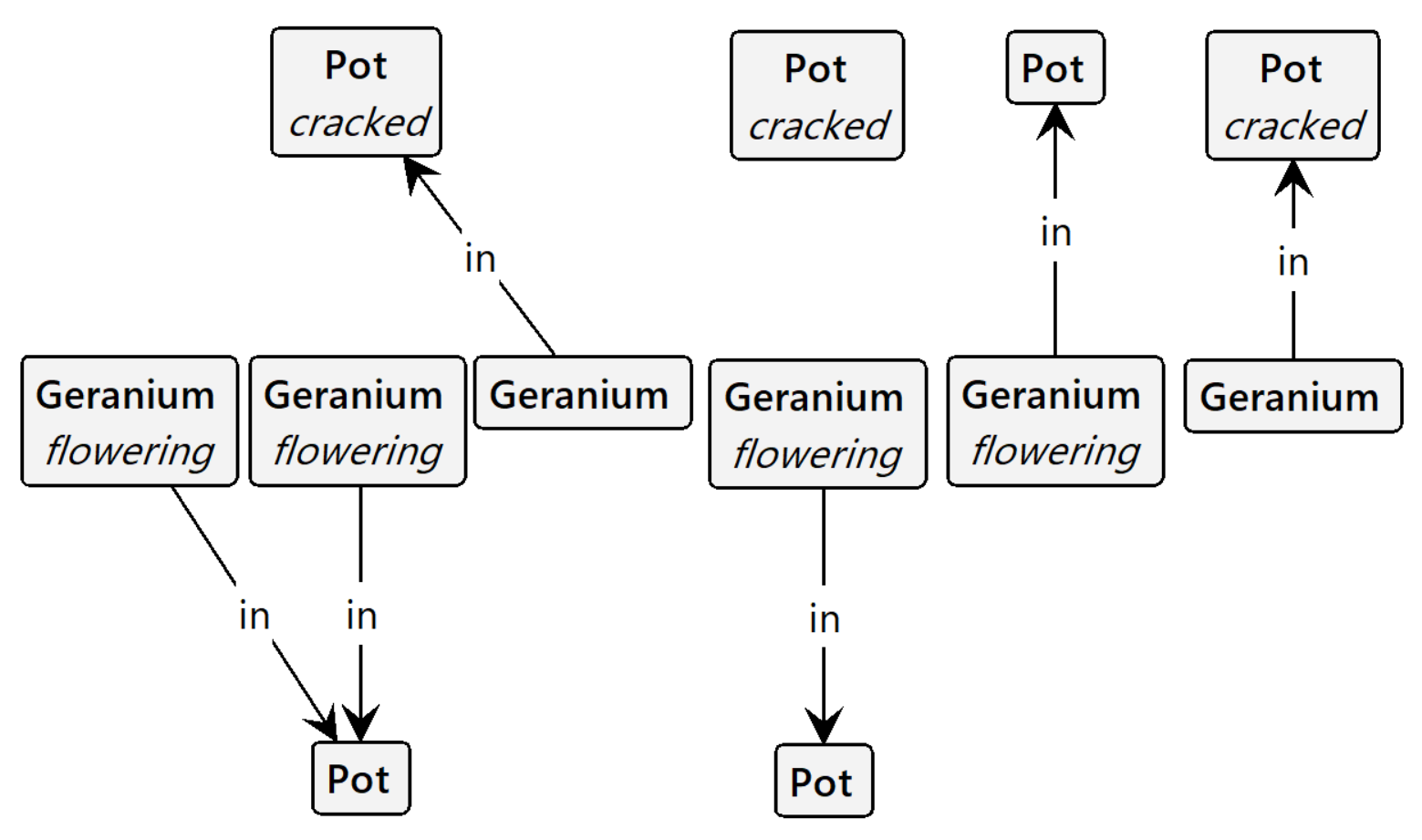}
	\subcaption{Graph after rewriting}
	\end{minipage}
	\caption{\newsentence{Repotting the Geraniums (screenshot from
          the GROOVE tool)}}
	\label{fig:Repotting the Geraniums in GROOVE}
\end{figure}

\begin{Verbatim}[frame=single]
   M<+>( {type(pot),flag(cracked),N<+>t(X),$p},
       N<+>{type(geranium),flag(flowering),s(X)} ) :-
   M<+>( {type(pot),flag(cracked),$p}, {type(pot),N<+>t(X)},
       N<+>{type(geranium),flag(flowering),s(X)} )
\end{Verbatim}

\begin{figure}[t]
	\begin{minipage}[b]{0.48\linewidth}
	\centering
	\includegraphics[height=3.5cm]{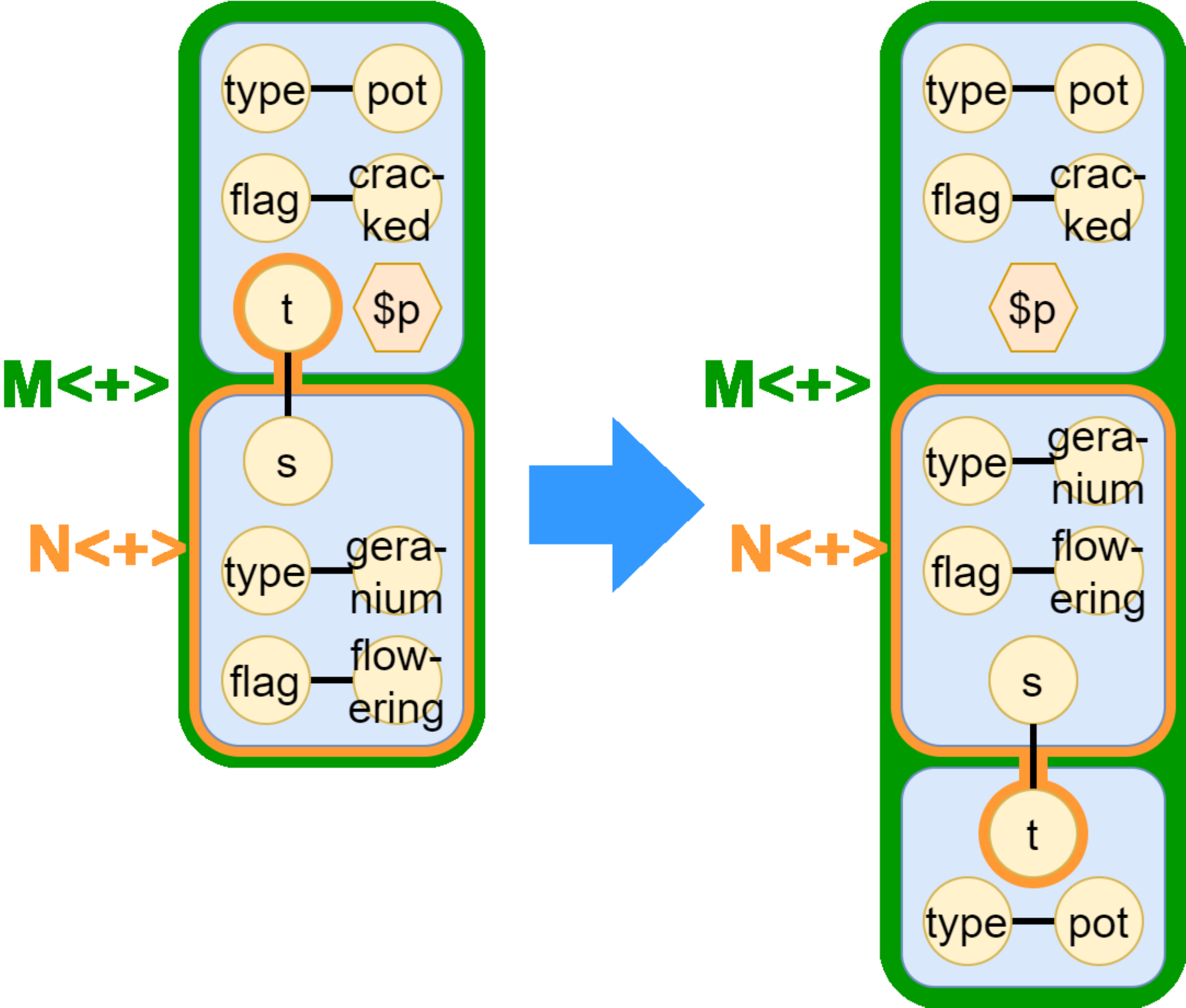}
	\subcaption{Rule using links}
	\end{minipage}
	\begin{minipage}[b]{0.48\linewidth}
	\centering
        \includegraphics[height=2cm]{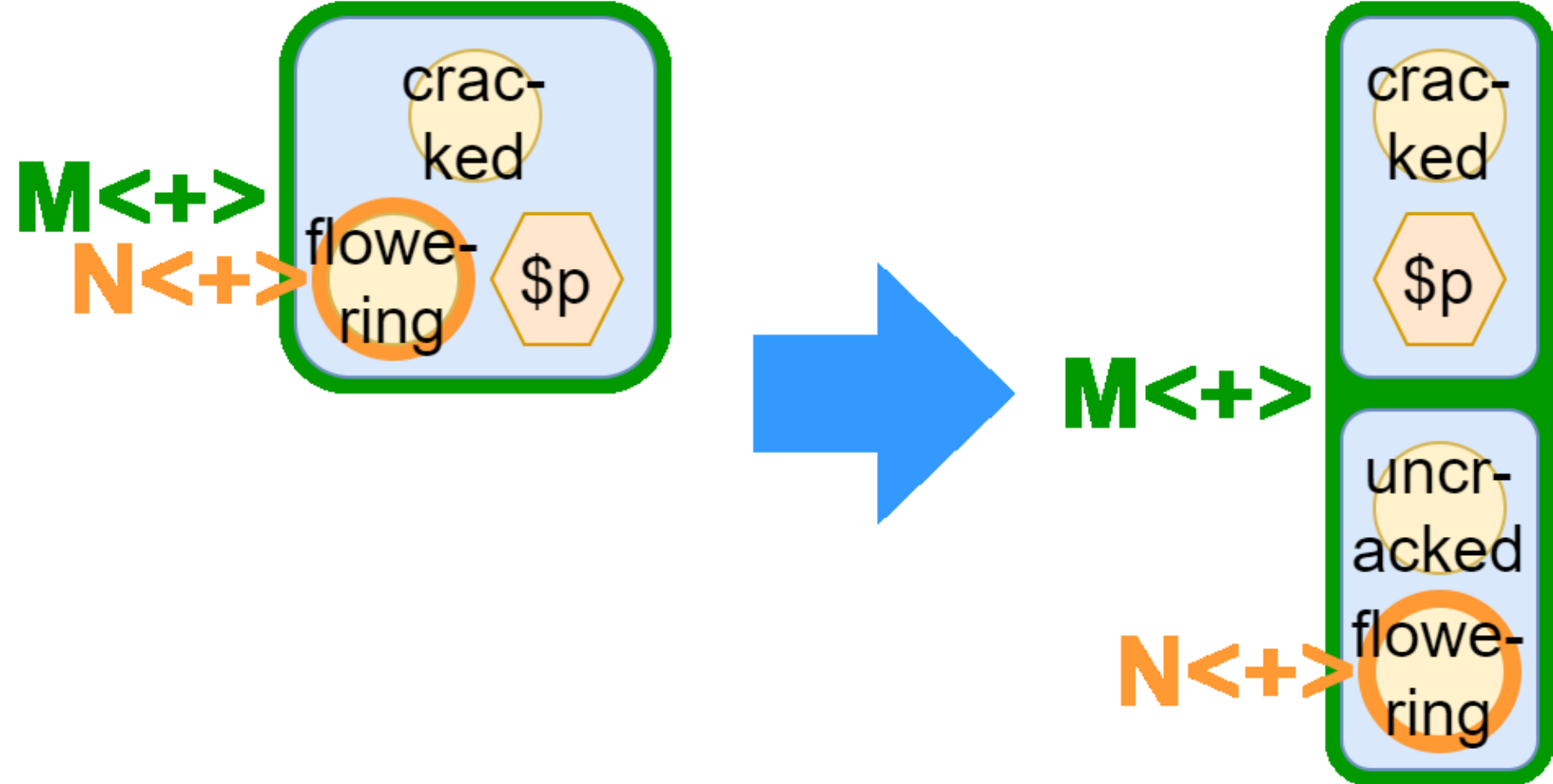}
	\vspace*{20pt}
	\subcaption{Rule without links}
	\end{minipage}
	\caption{\newsentence{Repotting the Geraniums in QLMNtal}}
	\label{fig:Repotting the Geraniums in QLMNtal}
\end{figure}


\newsentence{\noindent
A visual version of this rule is given in
Fig.~\ref{fig:Repotting the Geraniums in QLMNtal}(a), where
the scope of the two quantifiers are shown in two different colors.}

Each object is represented by a membrane, with the type (\texttt{pot}
or \texttt{geranium}) represented by a unary atom connected to a
\texttt{type} atom and the state (\texttt{cracked},
\texttt{flowering}) by a unary atom connected to a \texttt{flag} atom. 
The location of entities is represented by connecting membranes with
\texttt{s} (source) atoms and \texttt{t} (target) atoms.  
The rewrite rule uses nested universal quantifiers, where
the outer universal quantifier (labeled \texttt{M}) matches all
cracked pots 
and the inner universal quantifier (labeled \texttt{N})
matches all flowering geraniums in each cracked pot.  
Because of the `\texttt{<+>}' quantifier, 
no new pots are prepared for pots without a flowering geranium, even
if they are cracked.

\newsentence{%
Whereas the above rule is intended to reflect the original graph
rewriting rule \cite{Geraniums}, the membrane construct of LMNtal
allows more concise description of the problem that does not use
any links to represent placement:}

\begin{Verbatim}[frame=single]
   M<+>{cracked, N<+>flowering, $p} :-
   M<+>({cracked, $p}, {uncracked, N<+>flowering}).
\end{Verbatim}

\noindent
\newsentence{A visual version of this simplified rule is shown in 
Fig.~\ref{fig:Repotting the Geraniums in QLMNtal}(b).}


\subsection{Petri Nets}
\label{subsec:Petri}

\begin{wrapfigure}{r}{0.4\textwidth}
\vspace*{-\intextsep}
	\begin{minipage}[b]{0.41\linewidth}
		\centering
		\includegraphics[width=2.0cm]{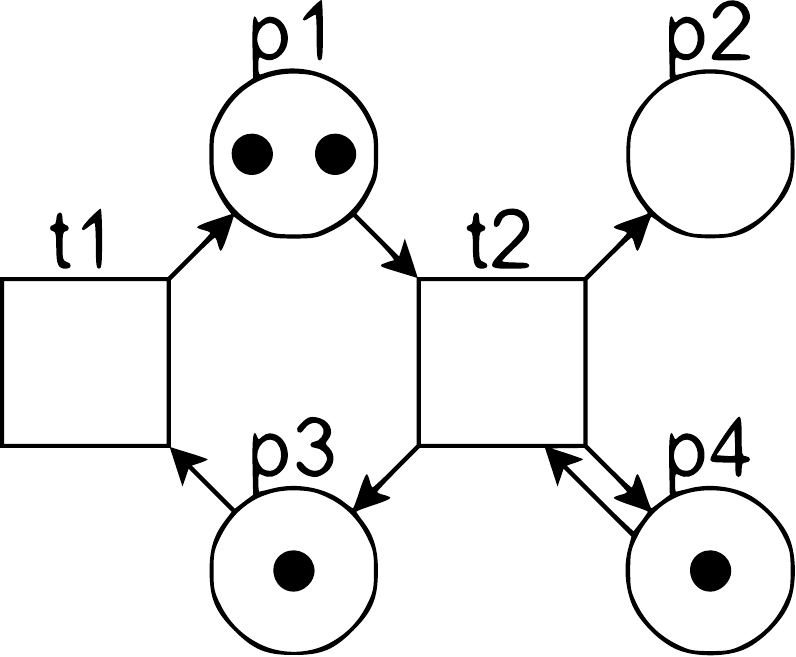}
	\end{minipage}
\quad
	\begin{minipage}[b]{0.41\linewidth}
		\centering
		\includegraphics[width=2.0cm]{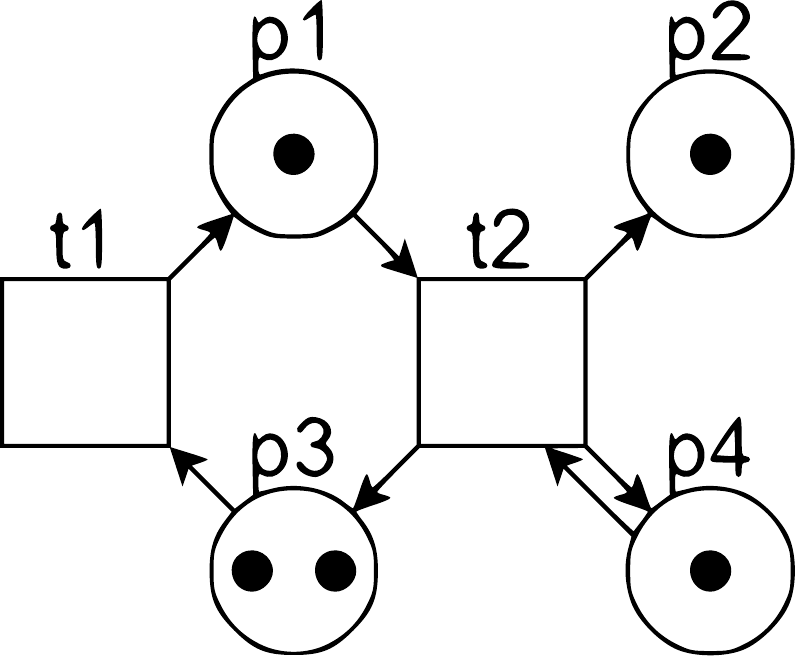}
	\end{minipage}
	\caption{Petri Net, before and after transition}
	\label{fig:Petri Net}
\vspace*{-16pt}
\end{wrapfigure}

\begin{figure}[t]
	\centering
	\includegraphics[height=1.3cm]{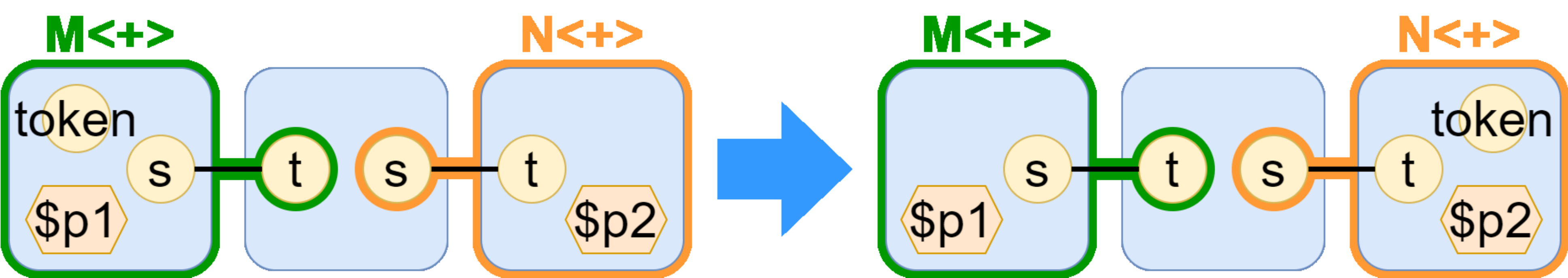}
	\caption{\newsentence{Petri Net in QLMNtal}}
	\label{fig:Petri Net in QLMNtal}
\end{figure}

Petri Nets \cite{PetriNet} describe discrete event systems as
directed bipartite graphs.
The most fundamental form, place-transition nets, consists of places
(circles), transitions (rectangles), arcs and tokens (bullets).
A transition fires when every input place of the transition has at
least one token.  When the transition fires, it
removes one token from each of input places and adds
one token to every output place.
Figure~\ref{fig:Petri Net} shows how \texttt{t2} fires
(both \texttt{t1} and \texttt{t2} may fire).
This can be encoded concisely in QLMNtal as follows,

\begin{Verbatim}[frame=single]
  M<+>{s(A1),token,$1}, {M<+>t(A1),N<+>s(A2)}, N<+>{t(A2),$2} :-
  M<+>{s(A1),$1}, {M<+>t(A1),N<+>s(A2)}, N<+>{t(A2),token,$2}
\end{Verbatim}

\noindent
where the left and the right membranes in the head and the body
represent places, while the middle membrane represents transition.
We used labels \texttt{M} and \texttt{N} to indicate which
quantifiers are interdependent and should be unrolled together.
\newsentence{Figure~\ref{fig:Petri Net in QLMNtal}
shows a visual representation of the QLMNtal rule.}

%% file: tex/sec7.tex
\section{Related and Future Work}
\label{sec:related}

We first note that, while there are a number of graph rewriting tools
\cite{VIATRA2,GrGen,GReAT,PROGRES,GP2,GROOVE,PORGY},
many of them are visual tools, and our work can be regarded as an
attempt to design a \textit{language with
abstract syntax} and \textit{syntax-directed operational semantics}
as in other
programming languages.  We share this motivation with GP 2 \cite{GP2},
but the design principles of the two languages are very different.

\newsentence{%
Most of the graph rewriting tools provide \textit{control structure} for
rewriting,
some in visual syntax (e.g., GReAT \cite{GReAT}) and some in textual syntax
(e.g., the strategy language of PORGY \cite{PORGY}).
In contrast to all these tools that come with separate sublanguages for
execution control, (Q)LMNtal is designed as an inherently concurrent
language whose execution is controlled by subgraph matching that serves
as a synchronization mechanism.}

While many graph rewriting tools support 
non-existence conditions
\cite{VIATRA2,GrGen,GReAT,PROGRES},
fewer tools feature
universal quantification within a rewrite rule.
%
%
The goal of 
QLMNtal is to enable all quantifications to be expressed within a single 
rewrite rule, 
thereby simplifying the state space.

The notion of cardinality can be found in (i) practical 
regular expressions, (ii) answer set programming (ASP) \cite{ASP} for
satisfiability and optimization problems, and (iii) some graph 
formalisms \cite{QGPs,GReAT,QGRAPH}.
QLMNtal can not only specify how many copies of a subgraph should be 
extracted but can also rewrite all the extracted graphs
in a single step.

An important feature of GROOVE \cite{GROOVE} is that it allows
combined and nested use of quantification within a single rule.
Non-existence is expressed by coloring the graph, 
while existential and universal quantification are expressed by
connecting them to special nodes representing quantifiers. 
In QLMNtal, the 
syntax-directed
operational semantics
allows various quantifications (including non-existence) to be
combined and nested, where labeled quantifiers specify which elements
of rewrite rules are jointly quantified.
We believe our operational semantics can lead smoothly to 
logical interpretation of the proposed constructs, which is our future work.

There have been several attempts to support graph rewriting in
functional languages \cite{Clean,FUnCAL,LambdaGT}.
However, in functional graph rewriting languages, 
handling of quantities seems (to the best of our knowledge) to be an
open problem.

%

Future work includes extending SLIM and the LMNtal compiler. 
SLIM is a
virtual machine for LMNtal with dedicated intermediate code.
%
Our previous work \cite{Saito} showed that 
a limited form of nested quantification
could be checked by SLIM's instruction set.
Our 
challenge is to support the
full expressive power of QLMNtal by introducing additional
control structure into SLIM that an enhanced LMNtal compiler
should then support.


%% file: tex/ack.tex
\subsubsection*{Acknowledgments}
The authors are indebted to anonymous reviewers for their valuable
comments. 
This work is partially supported by Grant-In-Aid for Scientific
Research (23K11057), JSPS, Japan, and Waseda University Grant for
Special Research Projects (2024C-432).

%% file: tex/appendix.tex
\appendix
\section{Appendix}
\label{sec:appendix}

\subsection{Examples with explanations of the semantics}
\label{sec:appendix_explanation}

%
For the design validation of the semantics with (EQ) and (RQ)
(Section~5), i.e., to see that the semantics
reflects the intended behavior, we explain how it works using
several examples.
%
The rewriting can be subdivided into the following steps:

\begin{enumerate}
	\item \textit{Expanding Abbreviations}
	\item \textit{Application of }(EQ)
	\item \textit{Checking for }(RQ$'$)
	\item \textit{Checking for }(\textit{CardCond})
	\item \textit{Checking for }(\textit{NegCond})
	\item \textit{Application of }(RQ) 
                      possibly with the help of (R3$'$)
\end{enumerate}

In Step 1, the rules are transformed into their unabbreviated form.

In Step 2, one or more cases are generated by applying (EQ) to the
rules.
If the rules contain cardinality quantification templates, 
the result of Steps 3--5 depends on how many copies are generated by
(EQ) from the cardinality quantified processes.
Note that (EQ) itself may generate infinitely many cases (including
those that do not satisfy (\textit{CardCond})),
but in this section we focus on a few cases. 

In Steps 3--5, we check if (RQ) is applicable. 
In Step 3, we apply $\textit{simp}$ to the heads and the bodies of the
rules 
and check if (RQ$'$) is applicable to the program. 
In Step 4, we check if all the outermost cardinality quantifiers
appearing in the rule satisfy (\textit{CardCond}).  
In Step 5, we check if all the outermost non-existence quantifiers
appearing in the rule satisfy (\textit{NegCond}).  

In Step 6, we actually rewrite the process by applying (RQ), which was confirmed to be applicable in Steps 3--5.

\subsubsection{Program 1:}

Rewrite all unary \texttt{a}'s
to \texttt{c}'s and all unary \texttt{b}'s to \texttt{d}'s.

\medskip
\begin{screen}[4]
\begin{center}
\begin{tabular}{rl}
           Rule: $R_1 =\ $& \texttt{M<*>a(X),N<*>b(Y) :- M<*>c(X),N<*>d(Y).}\\[2pt]
Initial Process: $P_1 =\ $& \texttt{a(X),b(X),b(Y),c(Y).}
\end{tabular}
\end{center}
\end{screen}

\step{Expanding Abbreviations}{}:

\begin{center}
\begin{tabular}{rl}
           Rule: $R_1 =\ $& \texttt{\underline{\card{M}{0}{$\infty$}{a(X)},M<\caret>a(V)},\underline{\card{N}{0}{$\infty$}{b(Y)},N<\caret>b(W)}}\\
		                & \texttt{:- \underline{\card{M}{0}{$\infty$}{c(X)},M<\caret>c(V)},\underline{\card{N}{0}{$\infty$}{d(Y)},N<\caret>d(W)}.}\\
\end{tabular}
\end{center}

\step{Application of}{(EQ):} 
(the underlined atoms have been replicated)

\medskip\noindent
\begin{tabular}{cl}
& \texttt{\card{M}{0}{$\infty$}{a(X)},M<\caret>a(V),\card{N}{0}{$\infty$}{b(Y)},N<\caret>b(W)}\\
& \texttt{:- \card{M}{0}{$\infty$}{c(X)},M<\caret>c(V),\card{N}{0}{$\infty$}{d(Y)},N<\caret>d(W)}\\[-2pt]
$\stackrel{\textrm{(EQ)}}{\equiv}$ &
\texttt{\card{M}{-1}{$\infty$}{a(X)},\underline{a(X1)},M<\caret>a(V),\card{N}{0}{$\infty$}{b(Y)},N<\caret>b(W)}\\
& \texttt{:- \card{M}{-1}{$\infty$}{c(X)},\underline{c(X1)},M<\caret>c(V),\card{N}{0}{$\infty$}{d(Y)},N<\caret>d(W)}\hspace{14pt}(\textit{Case 1})\\[-2pt]
$\stackrel{\textrm{(EQ)}}{\equiv}$ &
\texttt{\card{M}{-1}{$\infty$}{a(X)},a(X1),M<\caret>a(V),}%
\texttt{\card{N}{-2}{$\infty$}{b(Y)},\underline{b(Y1),b(Y2)},N<\caret>b(W)}\\
&\texttt{:-}\\
&\texttt{\card{M}{-1}{$\infty$}{c(X)},c(X1),M<\caret>c(V),\card{N}{-2}{$\infty$}{d(Y)},\underline{d(Y1),d(Y2)},N<\caret>d(W)}\\
\multicolumn{2}{r}{(\textit{Case 2})}
\end{tabular}

\medskip\noindent
\textbf{\textit{Case 1}}

\begin{screen}[4]


\begin{center}
\begin{tabular}{rl}
  Rule: $R_1^1 =\ $& \texttt{\card{M}{-1}{$\infty$}{a(X)},a(X1),M<\caret>a(V),}\\
  &\texttt{\card{N}{0}{$\infty$}{b(Y)},N<\caret>b(W)}\\
  &\texttt{:-}\\
  &\texttt{\card{M}{-1}{$\infty$}{c(X)},c(X1),M<\caret>c(V),}\\
  &\texttt{\card{N}{0}{$\infty$}{d(Y)},N<\caret>d(W).}\\[2pt]
Initial Process: $P_1 =\ $& \texttt{a(X),b(X),b(Y),c(Y).}
\end{tabular}
\end{center}
\end{screen}

\step{Checking for}{(RQ$'$)}, where
%
%
%
$\theta=[(\texttt{b(X),b(Y),c(Y)})/\texttt{\$}\self]$:
%
%
%
$$\begin{tabular}{rl}
&\texttt{\underline{a(X)},b(X),b(Y),c(Y)}\\
&($\equiv$ $(\func(\texttt{\card{M}{-1}{$\infty$}{a(X)},a(X1),M<\caret>a(V),\card{N}{0}{$\infty$}{b(Y)},N<\caret>b(W)$)$,\$$\self$})\theta$\;)\\
$\reducesR{R_1^1}$&
\texttt{\underline{c(X)},b(X),b(Y),c(Y)}\\
&($\equiv$ $(\func(\texttt{\card{M}{-1}{$\infty$}{c(X)},c(X1),M<\caret>c(V),\card{N}{0}{$\infty$}{d(Y)},N<\caret>d(W)$)$,\$$\self$})\theta$\;)
\end{tabular}$$

\step{Checking for}{(\textit{CardCond})}:

(\textit{CardCond}) for the \texttt{\card{M}{-1}{$\infty$}{}}: $-1\le
0\;\land\;\infty\ge 0$ $~ = \textit{true}$ 

(\textit{CardCond}) for the \texttt{\card{N}{0}{$\infty$}{}}:  $0\le
0\;\land\;\infty\ge 0$ $~ = \textit{true}$ 

\step{Checking for}{(\textit{NegCond})}:

(\textit{NegCond}) for the \texttt{M<\caret>}: 
%
$\texttt{\{b(X),b(Y),c(Y),\ttnameid{$\self$}\}}
\hspace{40pt}{\not}\hspace{-40pt}
\reducesR{\raise2pt\hbox{\scriptsize$
\texttt{\{a(X),\$\ttnamecxt{$\self$},\ttnameid{$\self$}\}}\react$}}$ 
%
%
$~ =\textit{true}$

(\textit{NegCond}) for the \texttt{N<\caret>}: 
%
$\texttt{\{\underline{b(X)},b(Y),c(Y),\ttnameid{$\self$}\}}
\hspace{40pt}{\not}\hspace{-40pt}
\reducesR{\raise2pt\hbox{\scriptsize$
\texttt{\{b(Y),\$\ttnamecxt{$\self$},\ttnameid{$\self$}\}}\react$}}$ 
%
%
$~ =\textit{false}$

\medskip\noindent
In Case 1, rewriting is not possible
because \texttt{N<\caret>} does not satisfy (\textit{NegCond}).

\medskip\noindent
\textbf{\textit{Case 2}}

\begin{screen}[4]


	\begin{center}
	\begin{tabular}{rl}
	Rule: $R_1^2 =\ $& \texttt{\card{M}{-1}{$\infty$}{a(X)},a(X1),M<\caret>a(V),}\\
&\texttt{\card{N}{-2}{$\infty$}{b(Y)},b(Y1),b(Y2),N<\caret>b(W)}\\
	&\texttt{:-}\\
        &\texttt{\card{M}{-1}{$\infty$}{c(X)},c(X1),M<\caret>c(V),}\\
    &\texttt{\card{N}{-2}{$\infty$}{d(Y)},d(Y1),d(Y2),N<\caret>d(W).}\\[2pt]
	Initial Process: $P_1 =\ $& \texttt{a(X),b(X),b(Y),c(Y).}
	\end{tabular}
	\end{center}
\end{screen}

\step{Checking for}{(RQ$'$)}, where
%
%
$\theta=[(\texttt{c(Y)})/\texttt{\$}\self]$:
%
%
%
{
$$\begin{tabular}{rl}
&\texttt{\underline{a(X),b(X),b(Y)},c(Y)}\\
&($\equiv$ $(\func(\texttt{\card{M}{-1}{$\infty$}{a(X)},a(X1),M<\caret>a(V),}$\\
&$\qquad\qquad\ \texttt{\card{N}{-2}{$\infty$}{b(Y)},b(Y1),b(Y2),N<\caret>b(W)$)$,\$$\self$})\theta$\;)\\
$\reducesR{R_1^2}$&
\texttt{\underline{c(X),d(X),d(Y)},c(Y)}\\
&($\equiv$ $(\func(\texttt{\card{M}{-1}{$\infty$}{c(X)},c(X1),M<\caret>c(V),}$\\
&$\qquad\qquad\ \texttt{\card{N}{-2}{$\infty$}{d(Y)},d(Y1),d(Y2),N<\caret>d(W)$)$,\$$\self$})\theta$\;)
\end{tabular}$$
}

\step{Checking for}{(\textit{CardCond})}:

(\textit{CardCond}) for the \texttt{\card{M}{-1}{$\infty$}{}}: $-1\le 0\;\land\;\infty\ge 0$ $~ = \textit{true}$

(\textit{CardCond}) for the \texttt{\card{N}{-2}{$\infty$}{}}: $-2\le 0\;\land\;\infty\ge 0$ $~ = \textit{true}$

\step{Checking for}{(\textit{NegCond})}:

(\textit{NegCond}) for the \texttt{M<\caret>}: 
%
$\texttt{\{c(Y),\ttnameid{$\self$}\}}
\hspace{40pt}{\not}\hspace{-40pt}
\reducesR{\raise2pt\hbox{\scriptsize$
\texttt{\{a(X),\$\ttnamecxt{$\self$},\ttnameid{$\self$}\}}\react$}}$ 
%
%
$~ =\textit{true}$

(\textit{NegCond})
for the \texttt{N<\caret>}: 
%
$\texttt{\{c(Y),\ttnameid{$\self$}\}}
\hspace{40pt}{\not}\hspace{-40pt}
\reducesR{\raise2pt\hbox{\scriptsize$
\texttt{\{b(X),\$\ttnamecxt{$\self$},\ttnameid{$\self$}\}}\react$}}$ 
%
%
$~ =\textit{true}$

\step{Application of}{(RQ)}:

\texttt{a(X),b(X),b(Y),c(Y)}
$\reducesR{\raise2pt\hbox{\scriptsize
\texttt{M<*>a(X),N<*>b(Y) :- M<*>c(X),N<*>d(Y)}}}$\\
\indent
\texttt{c(X),d(X),d(Y),c(Y)}

\subsubsection{Program 2:}

Rewrite 2 or 4 \texttt{a}'s to the same
number of \texttt{b}'s.

\medskip
\begin{screen}[4]
	\begin{center}
	\begin{tabular}{rl}
			   Rule: $R_2 =\ $& \texttt{\card{}{1}{2}{\num{}{2}{a}} :- \card{}{1}{2}{\num{}{2}{b}}.}\\[2pt]
	Initial Process: $P_2 =\ $& \texttt{a,a,a,a.}
	\end{tabular}
	\end{center}
\end{screen}

\step{Expanding Abbreviations}{}:


\begin{center}
\begin{tabular}{rl}
           Rule: $R_2 =\ $& \texttt{\card{}{1}{2}{\underline{\card{}{2}{2}{a}}} :- \card{}{1}{2}{\underline{\card{}{2}{2}{b}}}}
\end{tabular}
\end{center}

\step{Application of}{(EQ)}:

\medskip{
\leavevmode
\hbox to0pt{\hspace*{285pt}\mbox{(\textit{Case 1})}\hss}%
\hspace{10pt}\texttt{\card{}{1}{2}{\card{}{2}{2}{a}} :- \card{}{1}{2}{\card{}{2}{2}{b}}}
}

\noindent{
$\stackrel{\textrm{(EQ)}}{\equiv}$
\hbox to0pt{}
\texttt{\card{}{0}{1}{\card{}{2}{2}{a}},\underline{\card{M1}{2}{2}{a}} :- \card{}{0}{1}{\card{}{2}{2}{b}},\underline{\card{M1}{2}{2}{b}}}}

\noindent{
$\stackrel{\textrm{(EQ)}}{\equiv}$
\hbox to0pt{\hspace*{280pt}\mbox{(\textit{Case 2})}\hss}
\texttt{\card{}{0}{1}{\card{}{2}{2}{a}},\card{M1}{0}{0}{a},\underline{a,a} :- \card{}{0}{1}{\card{}{2}{2}{b}},\card{M1}{0}{0}{b},\underline{b,b}}}

\noindent{
$\stackrel{\textrm{(EQ)}}{\equiv}$
\hbox to0pt{}
\texttt{\card{}{-1}{0}{\card{}{2}{2}{a}},\card{M1}{0}{0}{a},a,a,\underline{\card{M2}{2}{2}{a}}}\\
\null\texttt{~~~~~~:- \card{}{-1}{0}{\card{}{2}{2}{b}},\card{M1}{0}{0}{b},b,b,\underline{\card{M2}{2}{2}{b}}}
}

\noindent
{
$\stackrel{\textrm{(EQ)}}{\equiv}$
\texttt{\card{}{-1}{0}{\card{}{2}{2}{a}},\card{M1}{0}{0}{a},a,a,\card{M2}{0}{0}{a},\underline{a,a}}\\
\hbox to0pt{\hspace*{300pt}\mbox{(\textit{Case 3})}\hss}
\null\texttt{~~~~~~:- \card{}{-1}{0}{\card{}{2}{2}{b}},\card{M1}{0}{0}{b},b,b,\card{M2}{0}{0}{b},\underline{b,b}}
}

\medskip\noindent
\textbf{\textit{Case 1}}

\begin{screen}[4]
	\begin{center}
	\begin{tabular}{rl}
	   Rule: $R_2^1 =\ $& \texttt{\card{}{1}{2}{\card{}{2}{2}{a}} :- \card{}{1}{2}{\card{}{2}{2}{b}}.}\\[2pt]
	Initial Process: $P_2 =\ $& \texttt{a,a,a,a.}
	\end{tabular}
	\end{center}
\end{screen}

\step{Checking for}{(RQ$'$)}, where
%
%
%
$\theta=[(\texttt{a,a,a,a})/\texttt{\$}\self]$:



$$\begin{tabular}{rl}
&\texttt{a,a,a,a}\
($\equiv$ $(\func(\texttt{\card{}{1}{2}{\card{}{2}{2}{a}}$)$,\$$\self$})\theta$\;)\\
$\reducesR{R_2^1}$&
\texttt{a,a,a,a}\
($\equiv$ $(\func(\texttt{\card{}{1}{2}{\card{}{2}{2}{b}}$)$,\$$\self$})\theta$\;)
\end{tabular}$$

\step{Checking for}{(\textit{CardCond})}:

(\textit{CardCond}) for the \texttt{\card{}{1}{2}{}}: $\underline{1\le 0}\;\land\;2\ge 0$ $~ = \textit{false}$

\medskip\noindent
In Case 1, rewriting is not possible because \texttt{\card{}{1}{2}{}} does not satisfy (\textit{CardCond}).

\medskip\noindent
\textbf{\textit{Case 2}}

\begin{screen}[4]
	\begin{center}
	\begin{tabular}{rl}
	   Rule: $R_2^2 =\ $& \texttt{\card{}{0}{1}{\card{}{2}{2}{a}},\card{M1}{0}{0}{a},a,a}\\
                & \texttt{:- \card{}{0}{1}{\card{}{2}{2}{b}},\card{M1}{0}{0}{b},b,b.}\\[2pt]
	Initial Process: $P_2 =\ $& \texttt{a,a,a,a.}
	\end{tabular}
	\end{center}
\end{screen}

\step{Checking for}{(RQ$'$)}, where
%
%
%
$\theta=[(\texttt{a,a})/\texttt{\$}\self]$:



$$\begin{tabular}{rl}
&\texttt{\underline{a,a},a,a}\
($\equiv$ $(\func(\texttt{\card{}{0}{1}{\card{}{2}{2}{a}},\card{M1}{0}{0}{a},a,a$)$,\$$\self$})\theta$\;)\\
$\reducesR{R_2^2}$&
\texttt{\underline{b,b},a,a}\
($\equiv$ $(\func(\texttt{\card{}{0}{1}{\card{}{2}{2}{b}},\card{M1}{0}{0}{b},b,b$)$,\$$\self$})\theta$\;)
\end{tabular}$$

\step{Checking for}{(\textit{CardCond})}:

(\textit{CardCond}) for the \texttt{\card{}{0}{1}{}}: $0\le 0\;\land\;2\ge 0$ $~ = \textit{true}$

(\textit{CardCond}) for the \texttt{\card{M1}{0}{0}{}}: $0\le 0\;\land\;0\ge 0$ $~ = \textit{true}$

\step{Checking for}{(\textit{NegCond})}:

Negation quantifier does not occur in this rule.

\step{Application of}{(RQ)}:

\begin{center}
\texttt{a,a,a,a}
$\reducesR{\raise2pt\hbox{\scriptsize
\texttt{\card{}{1}{2}{\num{}{2}{a}} :- \card{}{1}{2}{\num{}{2}{b}}}}}$
\texttt{b,b,a,a}
\end{center}

\medskip\noindent
\textbf{\textit{Case 3}}

\begin{screen}[4]


	\begin{center}
	\begin{tabular}{rl}
	   Rule: $R_2^3 =\ $& \texttt{\card{}{-1}{0}{\card{}{2}{2}{a}},\card{M1}{0}{0}{a},a,a,\card{M2}{0}{0}{a},a,a}\\
                & \texttt{ :- \card{}{-1}{0}{\card{}{2}{2}{b}},\card{M1}{0}{0}{b},b,b,\card{M2}{0}{0}{b},b,b.}\\[2pt]
	Initial Process: $P_2 =\ $& \texttt{a,a,a,a.}
	\end{tabular}
	\end{center}
\end{screen}

\step{Checking for}{(RQ$'$)}, where
%
%
%
$\theta=[\textbf{0}/\texttt{\$}\self]$:



$$\begin{tabular}{rl}
&\texttt{\underline{a,a,a,a}}\
($\equiv$ $(\func(\texttt{\card{}{-1}{0}{\card{}{2}{2}{a}},\card{M1}{0}{0}{a},a,a,\card{M2}{0}{0}{a},a,a$)$,\$$\self$})\theta$\;)\\
$\reducesR{R_2^3}$&
\texttt{\underline{b,b,b,b}}\
($\equiv$ $(\func(\texttt{\card{}{-1}{0}{\card{}{2}{2}{b}},\card{M1}{0}{0}{b},b,b,\card{M2}{0}{0}{b},b,b$)$,\$$\self$})\theta$\;)
\end{tabular}$$

\step{Checking for}{(\textit{CardCond})}:

(\textit{CardCond}) for the \texttt{\card{}{0}{1}{}}:  $0\le 0\;\land\;1\ge 0$ $~ = \textit{true}$

(\textit{CardCond}) for the \texttt{\card{M1}{0}{0}{}}: $0\le 0\;\land\;0\ge 0$ $~ = \textit{true}$

(\textit{CardCond}) for the \texttt{\card{M2}{0}{0}{}}: $0\le 0\;\land\;0\ge 0$ $~ = \textit{true}$

\step{Checking for}{(\textit{NegCond})}:

Negation quantifier does not occur in this rule.

\step{Application of}{(RQ)}

\begin{center}
\texttt{a,a,a,a}
$\reducesR{\raise2pt\hbox{\scriptsize
\texttt{\card{}{1}{2}{\num{}{2}{a}} :- \card{}{1}{2}{\num{}{2}{b}}}}}$
\texttt{b,b,b,b}
\end{center}

\subsubsection{Program 3:}

Generate an \texttt{ok} if there is no
membrane not containing an \texttt{a}.

\medskip
\begin{screen}[4]
	\begin{center}
	\begin{tabular}{rl}
	   Rule: $R_3 =\ $& \texttt{M<\caret>\{N<\caret>a,\$p\} :- ok.}\\[2pt]
	Initial Process: $P_3 =\ $& \texttt{\{a,b\}.}
	\end{tabular}
	\end{center}
\end{screen}

\medskip\noindent
Note that the two nested non-existence quantifiers are treated as
having different labels (Section \ref{sec:syntax}).

\step{Expanding Abbreviations}{}:


\begin{center}
\begin{tabular}{rl}
           Rule: $R_3 =\ $& \texttt{M<\caret>\{N<\caret>a,\$p\} :- \underline{M<\caret>\{N<\caret>a,\$p\}},ok}
\end{tabular}
\end{center}

\step{Checking for}{(RQ$'$)}, where
%
%
%
$\theta=[\texttt{\{a,b\}}/\texttt{\$}\self]$:



$$\begin{tabular}{rl}
&\texttt{\{a,b\}}\
($\equiv$ $(\func(\texttt{M<\caret>\{N<\caret>a,\$p\}$)$,\$$\self$})\theta$\;)\\
$\reducesR{R}$&
\texttt{\{a,b\},\underline{ok}}\
($\equiv$ $(\func(\texttt{M<\caret>\{N<\caret>a,\$p\},ok$)$,\$$\self$})\theta$\;)
\end{tabular}$$

\step{Checking for}{(\textit{CardCond})}:

Cardinality quantifier does not occur in this rule.

\step{Checking for}{(\textit{NegCond})}:

(\textit{NegCond}) for the \texttt{M<\caret>}: 
%
$\texttt{\{\{a,b,\ttnameid{p}\},\ttnameid{$\self$}\}}
\hspace{55pt}{\not}\hspace{-55pt}
\reducesR{\raise2pt\hbox{\scriptsize$\!\!
\texttt{\{\{N<\caret>a,\$\ttnamecxt{p},\ttnameid{p}\},\$\ttnamecxt{$\self$},\ttnameid{$\self$}\}}\react\!\!$}}%
~ =\textit{true}$ \\[-3pt]
%
%

\hspace{\fill}(Details of this is given
in ``\textbf{(\textit{NegCond})\textit{ for
the inner} \texttt{N<\caret>}}'' below) 

\step{Application of}{(RQ)}:

\begin{center}
\texttt{\{a,b\}}
$\reducesR{\raise2pt\hbox{\scriptsize
\texttt{M<\caret>\{N<\caret>a,\$p\} :- ok}}}$
\texttt{\{a,b\},ok}
\end{center}

\medskip\noindent
\textbf{\textit{(NegCond) for the inner} \texttt{N<\caret>}}

\begin{screen}[4]
	\begin{center}
	\begin{tabular}{rl}
	   Rule: $R'_3 =\ $& \texttt{\{\{N<\caret>a,\$\ttnamecxt{p},\ttnameid{p}\},\$$\itnamecxt{\self}$,$\itnameid{\self}$\} :- .}\\[2pt]
	Initial Process: $P'_3 =\ $& \texttt{\{\{a,b,\ttnameid{p}\},$\itnameid{\self}$\}}
	\end{tabular}
	\end{center}
\end{screen}

\step{Expanding Abbreviations}{}:


\begin{center}
\begin{tabular}{rl}
           Rule: $R'_3 =\ $& \texttt{\{\{N<\caret>a,\$\ttnamecxt{p},\ttnameid{p}\},\$$\itnamecxt{\self}$,$\itnameid{\self}$\} :- }\underline{\textbf{0}}
\end{tabular}
\end{center}

\step{Checking for}{(RQ$'$)}, where
%
%
%
$\theta=\bigl[(\texttt{a,b})/\texttt{\$\ttnamecxt{p}},\textbf{0}/\texttt{\$}\itnamecxt{\self},\textbf{0}/\texttt{\$}\self\bigr]$:
%
%
%
$$\begin{tabular}{rl}
&\texttt{\underline{\{\{a,b,\ttnameid{p}\},$\itnameid{\self}$\}}}\
($\equiv$ $(\func(\texttt{\{\{N<\caret>a,\$\ttnamecxt{p},\ttnameid{p}\},\$$\itnamecxt{\self}$,$\itnameid{\self}$\}$)$,\$$\self$})\theta$\;)\\
$\reducesR{R'_3}$&
\texttt{\underline{\textbf{0}}}\
($\equiv$ $(\func(\texttt{\textbf{0}$)$,\$$\self$})\theta$\;)
\end{tabular}$$

\step{Checking for}{(\textit{CardCond})}:

Cardinality quantifier does not occur in this rule.

\step{Checking for}{(\textit{NegCond})}:

(\textit{NegCond}) for the \texttt{N<\caret>}:\\[-4pt]
%
\indent
$\texttt{\{\{\{\underline{a},b,\ttnameid{$($\ttnamecxt{p}$)$}\},\ttnameid{$(\itnamecxt{\self})$}\},\ttnameid{$\self$}\}}
\hspace{95pt}{\not}\hspace{-95pt}$
$\reducesR{\raise2pt\hbox{\scriptsize$
\texttt{\{\{\{a,\$\ttnamecxt{$($\ttnamecxt{p}$)$},\ttnameid{$($\ttnamecxt{p}$)$}\},\$\ttnamecxt{$(\itnamecxt{\self})$},\ttnameid{$(\itnamecxt{\self})$}\},\$\ttnamecxt{$\self$},\ttnameid{$\self$}\}}\react$}}$\\[3pt]
%
%
\indent
$~ =\textit{false}$

\medskip\noindent
Rewriting with $R'_3$ is not done because \texttt{N<\caret>} 
does not satisfy (\textit{NegCond}).

\subsubsection{Program 4:}

Generate an \texttt{ok} if there is no
unordered pair of \texttt{a} and \texttt{b}.

\medskip
\begin{screen}[4]
	\begin{center}
	\begin{tabular}{rl}
	   Rule: $R_4 =\ $& \texttt{<\caret>a,<\caret>b :- ok.}\\[2pt]
	Initial Process: $P_4 =\ $& \texttt{a,a.}
	\end{tabular}
	\end{center}
\end{screen}

\step{Expanding Abbreviations}{}:


\begin{center}
\begin{tabular}{rl}
          Rule: $R_4 =\ $& \texttt{<\caret>a,<\caret>b :- ok}
\end{tabular}
\end{center}

\step{Checking for}{(RQ$'$)}, where
%
%
%
$\theta=[(\texttt{a,a})/\texttt{\$}\self]$:



$$\begin{tabular}{rl}
&\texttt{a,a}\
($\equiv$ $(\func(\texttt{<\caret>a,<\caret>b$)$,\$$\self$})\theta$\;)\\
$\reducesR{R_4}$&
\texttt{a,a,\underline{ok}}\
($\equiv$ $(\func(\texttt{ok$)$,\$$\self$})\theta$\;)
\end{tabular}$$

\step{Checking for}{(\textit{CardCond})}:

Cardinality quantifier does not occur in this rule.

\step{Checking for}{(\textit{NegCond})}:

(\textit{NegCond}) for the \texttt{<\caret>}: 
%
$\texttt{\{a,a,\ttnameid{$\self$}\}}
\hspace{35pt}{\not}\hspace{-35pt}
\reducesR{\raise2pt\hbox{\scriptsize$
\texttt{\{a,b,\$\ttnamecxt{$\self$},\ttnameid{$\self$}\}}\react$}}$ 
%
%
$~ =\textit{true}$

\step{Application of}{(RQ)}:

\begin{center}
\texttt{a,a}
$\reducesR{\raise2pt\hbox{\scriptsize
\texttt{<\caret>a,<\caret>b :- ok}}}$
\texttt{a,a,ok}
\end{center}

\subsubsection{Program 5:}

Generate an \texttt{ok} if there is
no \texttt{a} or \texttt{b}.

\medskip\noindent
Note the difference (from Program 4) of labeling to non-existence
quantifiers. 

\medskip
\begin{screen}[4]
	\begin{center}
	\begin{tabular}{rl}
	   Rule: $R_5 =\ $& \texttt{M<\caret>a,N<\caret>b :- ok.}\\[2pt]
	Initial Process: $P_5 =\ $& \texttt{a,a.}
	\end{tabular}
	\end{center}
\end{screen}

\step{Expanding Abbreviations}{}:


\begin{center}
\begin{tabular}{rl}
           Rule: $R_5 =\ $& \texttt{M<\caret>a,N<\caret>b :- ok}
\end{tabular}
\end{center}

\step{Checking for}{(RQ$'$)}, where
%
%
%
$\theta=[(\texttt{a,a})/\texttt{\$}\self]$:



$$\begin{tabular}{rl}
&\texttt{a,a}\
($\equiv$ $(\func(\texttt{M<\caret>a,N<\caret>b$)$,\$$\self$})\theta$\;)\\
$\reducesR{R_5}$&
\texttt{a,a,\underline{ok}}\
($\equiv$ $(\func(\texttt{ok$)$,\$$\self$})\theta$\;)
\end{tabular}$$

\step{Checking for}{(\textit{CardCond})}:

Cardinality quantifier does not occur in this rule.

\step{Checking for}{(\textit{NegCond})}:

(\textit{NegCond}) for the \texttt{M<\caret>}: 
%
$\texttt{\{\underline{a},a,\ttnameid{$\self$}\}}
\hspace{35pt}{\not}\hspace{-35pt}
\reducesR{\raise2pt\hbox{\scriptsize$
\texttt{\{a,\$\ttnamecxt{$\self$},\ttnameid{$\self$}\}}\react$}}$ 
%
%
$~ =\textit{false}$

(\textit{NegCond}) for the \texttt{N<\caret>}: 
%
$\texttt{\{a,a,\ttnameid{$\self$}\}}
\hspace{35pt}{\not}\hspace{-35pt}
\reducesR{\raise2pt\hbox{\scriptsize$
\texttt{\{b,\$\ttnamecxt{$\self$},\ttnameid{$\self$}\}}\react$}}$ 
%
%
$~ =\textit{true}$

\medskip\noindent
In this progam, rewriting is not done because \texttt{M<\caret>} does not satisfy (\textit{NegCond}).